\documentclass[11pt,a4paper]{article}
\usepackage{jheppub}
\usepackage{natbib}
\allowdisplaybreaks

\newcommand{\nablaslash}{\nabla{\!\!\!\!\slash}}

\newcommand{\bare}{\textrm{B}}
\newcommand{\tr}{\textrm{tr}}
\newcommand{\bphi}{\bar{\Phi}}
\newcommand{\bphid}{\bar{\Phi}^\dagger}

\begin{document}

\title{Yang-Mills Yukawa model in curved spacetime}

\author{David J. Toms}
\emailAdd{david.toms@newcastle.ac.uk}
\affiliation{Northumberland Centre for Theoretical Physics\footnote{current address}, Ladyshield, Haydon Bridge, Northumberland, U.K. NE47 6NH\\
and\\
School of Mathematics, Statistics and Physics,
Newcastle University,
Newcastle upon Tyne, U.K. NE1 7RU}

\date{\today}

\abstract
{The one-loop effective action for the scalar field part of a non-Abelian gauge theory based on a general gauge group of the form $G\times U(1)$, where the gauge group $G$ is arbitrary, is calculated. A complex scalar field, both Abelian and non-Abelian gauge fields and Dirac fermions coupled to gauge and scalar fields are included. A general mass term for the Dirac fields that includes a pseudoscalar term as well as both scalar and pseudoscalar Yukawa couplings is considered. The background field method is used in its manifestly gauge condition independent and gauge invariant form to isolate the divergent part of the one-loop effective action and to calculate the associated renormalisation group functions. Terms in the renormalised effective action up to and including those quadratic in the curvature are calculated using renormalisation group methods. The background scalar field is not assumed to be constant, so the second order derivative terms in the effective action can be calculated, and the gravitational background is kept arbitrary. The difference between the gauge condition independent approach and the more standard approach where gauge condition dependent results are found is demonstrated by explicit calculations. The anomaly that can arise if the pseudoscalar mass term for the fermions is transformed away from the classical theory is noted.}


\maketitle

\section{Introduction}\label{sec-intro}

The study of a general Yukawa model in curved spacetime was given recently~\cite{Tomsyukawa1,Tomsyukawa2}. This previous work considered only the case of an Abelian $U(1)$ gauge field. One of the main aims of the present paper is to generalise the calculations to non-Abelian theories. Our principle aim is to calculate the effective action for the scalar field part of a non-Abelian gauge theory based on a general gauge group of the form $G\times U(1)$, where the gauge group $G$ is arbitrary, at one-loop order. The compete form of the theory will be considered in Sec.~\ref{sec2} below. The calculations will be done by isolating the divergent parts of the effective action that just involve the scalar field (along with the vacuum part), so this paper is not a complete calculation of all of the one-loop divergences. The effective potential will then be calculated by using the renormalisation group as discussed originally in flat spacetime by Coleman and Weinberg \cite{ColemanandWeinberg} and generalised to curved spacetime by Buchbinder and Odintsov~\cite{BuchbinderandOdintsovRG} but with the neglect of the $R^2$ terms. (See the overview in \cite{ParkerTomsbook} based on earlier analysis in Ref.~\cite{TomsRG}.). The background field method of DeWitt~\citep{DeWittdynamical} will be utilised, but in the gauge invariant, gauge condition independent, and field parametrisation independent formulation given originally by Vilkovisky~\cite{Vilkovisky1}. The version of this applicable to any order in perturbation theory was given by DeWitt~\citep{DeWitt6}. The method is reviewed in \cite[Chap. 7]{ParkerTomsbook} and we will use the notation and conventions found there. We obtain a result for the effective action, including terms that involve the background scalar field gradient, at one-loop order up to and including terms that are quadratic in the curvature. This means that we do not have to restrict the background scalar field to be constant and do not assume that the background spacetime has constant curvature.

A selected set of earlier references to previous work on Yukawa interactions with scalars in curved spacetime includes \cite{shapiro1989asymptotic,odintsov1992general,elizalde1994renormalization,elizalde1995higgs, elizalde1995improved,elizalde1995renormalization,ProkopecWoodard,Garbrechtfermion,Garbrecht,Miao,Shapiro2011PLB, herranen2014spacetime,Cz,markkanen20181}. In particular, \cite{markkanen20181} has looked at the renormalisation group improved effective potential for the standard model in some detail and shown the potential importance of the $R^2$ terms in cosmology. Yukawa interactions have also been considered in related calculations: the asymptotic safety program for quantum gravity~\cite{Zanusso,Eichhornetal1,Eichhornetal3,oda2016non,Eichhornetal2}, in perturbative quantum gravity~\cite{RodigastSchuster}, in unimodular gravity~\cite{Martins2017PLB,Martins2018JCAP,Martins2018EPJC}, and in scale-invariant gravity~\cite{narain2017exorcising}. Some other recent work that is relevant includes \cite{jacobson2018nonperturbative} where non-perturbative effects in the effective action with Abelian gauge fields and Yukawa terms is considered; \cite{nakonieczny2019curved,barra2019renormalization} who who examine the role of Yukawa couplings in curved spacetime, but with no gauge fields.

The outline of our paper is as follows. The general model of a complex scalar field interacting with both Abelian and non-Abelian gauge fields as well as spinor fields is given in Sec.~\ref{sec2}. Both scalar and pseudoscalar mass terms are included, as are scalar and pseudoscalar Yukawa couplings. A brief description of the relevant parts of the background field method~\cite{DeWittdynamical,Vilkovisky1,DeWitt6,ParkerTomsbook} is given in Sec.~\ref{secbfmethod}, and the formal expression for the one-loop effective action is given. We present a discussion of the role and treatment of gauge conditions that is essential for understanding how the gauge condition independence is obtained later in the paper.  All of the pole parts for the one-loop effective action coming from the vector and scalar fields are found in Sec.~\ref{secBose}. The local momentum space method originated by Bunch and Parker~\cite{BunchParker} is used and some results are checked or obtained using heat kernel techniques. In Sec.~\ref{secspinors} we evaluate the pole part of the one-loop effective action that arises from the spinor fields by using results form our earlier work~\cite{Tomsyukawa1,Tomsyukawa2}. In Sec.~\ref{sec-ct} we combine all the results for the pole terms from previous sections and work out the necessary counterterms and renormalisation group functions. By keeping general parameters in the gauge conditions we show that all counterterms, apart from the field renormalisation factor, are independent of the gauge conditions. The correct gauge condition independent field renormalisation factor is obtained and it is emphasised how gauge condition dependent renormalisation group functions can result if this is not accounted for properly, and this has a knock-on effect for the potentials obtained in the subsequent section. In Sec.~\ref{effectivepotential} the renormalisation group is used to evaluate the effective potential and the gradient part of the effective action in the case where neither the scalar field nor the spinor fields have mass terms present. We present a short discussion in the last section and comment on an anomaly that was shown to arise~\cite{Tomsyukawa1} if the pseudoscalar mass terms for the fermions are removed by chiral transformations. Some of the technical calculations are contained or elaborated on in the Appendices.

\section{Non-Abelian gauged Yukawa model}\label{sec2}

We will consider a set of complex scalar fields arranged in a column vector $\Phi(x)$ coupled to an Abelian gauge field $A_\mu(x)$ and a non-Abelian gauge field $B_{A\mu}(x)$. We will use $A,B,\ldots$ to label the group indices for a simple gauge group $G$. The overall symmetry group will be $G\times U(1)$. We will not distinguish between upper and lower group indices and all repeated indices are summed over the dimension of $G$. Some useful results and conventions are summarised in Appendix~\ref{appendA}.

The bare action for the scalar field will be chosen to be
\begin{equation}
S_{\textrm{scalar}}=\int dv_x \Big\lbrack (D_\mu\Phi_{\textrm{B}})^\dagger(D^\mu\Phi_{\textrm{B}}) -m_{\textrm{s\,B}}^2\, |\Phi_{\textrm{B}}|^2 -\xi_{\textrm{B}}\,R \,|\Phi_{\textrm{B}}|^2 -\frac{\lambda_{\textrm{B}}}{6}\,|\Phi_{\textrm{B}}|^4\Big\rbrack,\label{2.1}
\end{equation}
where the subscript $B$ denotes a bare quantity. We take the spacetime to be curved and allow a non-minimal coupling to the curvature. $dv_x=|\det\,g_{\mu\nu}(x)|^{1/2}d^4x$ is the invariant spacetime volume element and $\nabla_\mu$ denotes the spacetime covariant derivative which for scalars is just the ordinary derivative. The gauge covariant derivative of $\Phi$ is defined by
\begin{equation}
D_\mu\Phi=\nabla_\mu\Phi-ie_s\,A_\mu\Phi-ig\,B_{A\mu}T_A\Phi,\label{2.2}
\end{equation}
with the scalar field $U(1)$ coupling constant $e_s$ and the non-Abelian gauge coupling constant $g$. The generators $T_A$ in \eqref{2.2} are in the appropriate representation for the scalar field and we choose these generators to be Hermitian. Because we are only considering the terms in the one-loop effective action that depend on a background scalar field we do not need to consider the renormalisation of either $e_s$ or $g$. This would not be the case if we were to consider the full renormalisation by including background gauge fields.

The vector field part of the action is
\begin{equation}
S_{\textrm{vector}}=-\frac{1}{4}\int dv_x \left(F^{\mu\nu}F_{\mu\nu}+B_{A\mu\nu}B_{A}{}^{\mu\nu}\right),\label{2.3}
\end{equation}
where the Abelian and non-Abelian field strengths are defined by
\begin{subequations}
\begin{align}
F_{\mu\nu}&=\nabla_\mu A_\nu-\nabla_\nu A_\mu,\label{2.3a}\\
B_{A\mu\nu}&=\nabla_\mu B_{A\nu}-\nabla_\nu B_{A\mu}+g f_{ABC}B_{B\mu}B_{C\nu},\label{2.3b}
\end{align}
\end{subequations}
respectively.

We also consider the coupling of the scalar field to spinor fields through a Yukawa coupling. We will allow for the presence of both scalar and pseudoscalar Yukawa couplings as well as scalar and pseudoscalar mass terms for the spinor fields. As a simple analogue of the $S(2)\times U(1)$ electroweak theory we will consider a $G$-singlet spinor field $\chi(x)$ that couples to the $U(1)$ vector field (as well as the scalar field), and another spinor field $\Psi(x)$ that transforms under $U(1)$ and under $G$ in the same way as the scalar field $\Phi(x)$. This allows for Yukawa interactions in the generic form $\chi\Phi^\dagger\Psi$ that we write in detail in \eqref{2.4}. The spinor field part of the action will be written as
\begin{align}
S_{\textrm{spinor}}&=\int dv_x\Big\lbrack \bar{\chi}(i\gamma^\mu D_\mu-m_{\chi}-im_{\chi5}\gamma_5)\chi+\bar{\Psi}(i\gamma^\mu D_\mu-m_{\psi}-im_{\psi5}\gamma_5)\Psi\nonumber\\
&\qquad\qquad-\bar{\chi}(w+iw_5\gamma_5)\Phi^\dagger\Psi-\bar{\Psi}\Phi(w^\ast+iw_5^\ast\gamma_5)\chi\Big\rbrack.\label{2.4}
\end{align}
Here
\begin{subequations}
\begin{align}
D_\mu\chi&=\nabla_\mu\chi-ie_\chi\,A_\mu \chi,\label{2.4a}\\
D_\mu\Psi&=\nabla_\mu\Psi-ie_\psi\,A_\mu \Psi-ig\,B_{A\mu}T_A \Psi,\label{2.4b}
\end{align}
\end{subequations}
define the gauge covariant derivatives with $e_\chi$ and $e_\psi$ the $U(1)$ coupling constants for $\chi$ and $\Psi$ respectively, and $g$ the coupling constant for $\Psi$ to the non-Abelian gauge field. The representation for $\Psi$ must be the same as that for $\Phi$ if the action in \eqref{2.4} is to make sense. In \eqref{2.4} we use $m_\chi$ and $m_{\chi5}$ to be the scalar and pseudoscalar mass terms for $\chi$ with similar expressions for $\Psi$. These mass terms could be generalised to non-trivial matrices but we will not consider this here. The scalar and pseudoscalar Yukawa couplings are denoted by $w$ and $w_5$ and are treated as complex numbers. The form of the action in \eqref{2.4} is real as written. It is customary to omit the pseudoscalar mass terms. In flat spacetime they can always be removed from the Dirac equation not coupled to any gauge fields by a chiral rotation of the spinor field. However in the presence of gravity or a non-trivial background gauge field an anomaly can arise~\cite{Tomsyukawa1,Tomsyukawa2}. These anomalies turn out not to affect the one-loop counterterms but we will keep them in and discuss them later in Sec.~\ref{effectivepotential}.

The $\gamma$-matrices in \eqref{2.4} are defined in terms of the usual flat spacetime $\gamma$-matrices as given by \cite{BjorkenandDrell1} for example by using the vierbein formalism. The covariant derivative $\nabla_\mu$ appearing in \eqref{2.4a} and \eqref{2.4b} contains the usual spin connection. We use the results and conventions of \cite[Chapter 5]{ParkerTomsbook} here. 

Because we are in curved spacetime we must also include the gravitational part of the action
\begin{equation}
S_{\textrm{grav}}=\int dv_x\Big(\Lambda_{\textrm{B}}+\kappa_{\textrm{B}}\,R+\alpha_{1\,\textrm{B}}\,R^{\mu\nu\lambda\sigma}R_{\mu\nu\lambda\sigma} + \alpha_{2\,\textrm{B}}\,R^{\mu\nu}R_{\mu\nu} + \alpha_{3\,\textrm{B}}\,R^2 \Big),\label{2.5}
\end{equation}
in order to remove divergences from the vacuum part of the effective action.

The complete classical action is then given by
\begin{equation}
S=S_{\textrm {scalar}}+S_{\textrm{vector}}+S_{\textrm{spinor}}+S_{\textrm{grav}}.\label{2.6}
\end{equation}
It will be invariant under the following infinitesimal gauge transformations
\begin{subequations}\label{2.7}
\begin{align}
\delta A_\mu&=\nabla_\mu\delta\epsilon,\label{2.7a}\\
\delta B_{A\mu}&=\nabla_\mu\delta\epsilon_A+g f_{ABC}B_{B\mu}\delta\epsilon_C,\label{2.7b}\\
\delta \Phi&=i\,(e_s\,\delta\epsilon+g\,\delta\epsilon_AT_A)\Phi,\label{2.7c}\\
\delta \Phi^\dagger&=-i\,\Phi^\dagger(e_s\,\delta\epsilon+g\,\delta\epsilon_AT_A),\label{2.7d}\\
\delta\chi&=i\,e_\chi\,\delta\epsilon\,\chi,\label{2.7e}\\
\delta \Psi&=i\,(e_\psi\,\delta\epsilon+g\,\delta\epsilon_AT_A)\Psi.\label{2.7f}
\end{align}
\end{subequations}
Here $\delta\epsilon$ is the infinitesimal parameter corresponding to the $U(1)$ part of the gauge group, and $\delta\epsilon_A$ are the infinitesimal parameters for the non-Abelian part $G$ of the gauge group.

In the next section we will consider the background field method applied to the scalar sector of the one-loop effective action based on \eqref{2.6}.

\section{Background field method and effective action}\label{secbfmethod}

The starting point is to expand all fields about a background part and a quantum part, where by quantum we mean that it is integrated over in the functional integral that defines the effective action. Because we only want the scalar field part of the effective action this means that we only need to consider background scalar fields; the background spinor and vector fields can be set equal to zero here. This prohibits us from performing a complete one-loop analysis of the counterterms but is sufficient for calculating the terms in the effective action that depend only on the background scalar fields. In particular we will not calculate the renormalisation group functions for the gauge coupling constants or the Yukawa coupling constants, but these are not required for the one-loop effective potential. Neglect of the background vector and spinor fields results in considerable simplification. Of course there is nothing stopping the approach being generalised to include the background vector and spinor fields other than technical complexity. 

We denote the background scalar field by $\bar{\Phi}$ and we do not assume that it is constant. This is essential if we are to calculate the relevant scalar field renormalisation. We replace $\Phi$ by $\bar{\Phi}+\Phi$ in \eqref{2.1} and \eqref{2.4} where $\Phi$ is now treated as a quantum field in the sense that it is integrated over in the functional integral. The gauge and spinor fields are treated as quantum. To one-loop order we just need terms in the action that are quadratic in the quantum fields. For the spinor part of the action \eqref{2.4}, because $\chi$ and $\Psi$ are treated as quantum fields, we simply take $\Phi=\bar{\Phi}$ to quadratic order. The quantum spinor fields do not couple to any of the other quantum fields so this enables us to consider the Bose and spinor contributions to the effective action separately as the functional integral will factorise. The spinor field contribution will be considered in Sec.~\ref{secspinors}. For now we concentrate on just the Bose fields. Our description here will be very brief and the reader unfamiliar with the formalism should consult \cite{Vilkovisky1,DeWitt6,ParkerTomsbook} for more details.

The vector part that is quadratic in the quantum fields is obtained from \eqref{2.3} by simply dropping the non-linear term in \eqref{2.3b}. The scalar field part coming from \eqref{2.1} is more complicated as it involves coupling terms between the gauge and scalar fields. Before writing this out we will consider the issue of gauge conditions.

If we use DeWitt's condensed notation~\cite{DeWittdynamical} and let $\varphi^i$ stand for all of the Bose fields, then the gauge transformations can be written as
\begin{equation}
\delta\varphi^i=K^{i}_{\alpha}\,\delta\epsilon^\alpha.\label{3.1}
\end{equation}
Here $\delta\epsilon^\alpha=(\delta\epsilon,\delta\epsilon_A)$ represents the complete set of infinitesimal gauge parameters, and the expressions for $K^i_\alpha$ follow from \eqref{2.7}. 

Central to the Vilkovisky-DeWitt construction of the effective action is the metric on the space of fields. The generators of the gauge transformations in \eqref{3.1} are Killing vectors for this metric. In the present theory the field space metric follows from the action \eqref{2.1} and \eqref{2.3} (by analogy with the non-linear sigma model) to have the non-zero components
\begin{subequations}\label{3.2}
\begin{align}
g_{\Phi_i(x)\Phi^\dagger_j(x')}=g_{\Phi^\dagger_i(x)\Phi_j(x')} &=\delta_{ij}\,\delta(x,x'), \label{3.2a}\\
g_{A_\mu(x)A_\nu(x')} &=-g^{\mu\nu}(x)\,\delta(x,x'), \label{3.2b}\\
g_{B_{A\mu}(x)B_{B\nu}(x')} &=-\delta_{AB}\,g^{\mu\nu}(x)\,\delta(x,x'). \label{3.2c}
\end{align}
\end{subequations}
(The minus signs that occur in \eqref{3.2b} and \eqref{3.2c} are a consequence of our choice for the spacetime metric signature.) In place of indices we use the actual fields to label the components. The utility of this was shown originally by Kunstatter~\cite{Kunstatter1}. An important point to note here is that because the field space metric has no explicit dependence on the fields the corresponding Christoffel connection will vanish.

The Landau-DeWitt gauge condition is defined by
\begin{equation}
C_\alpha=\bar{K}_{\alpha\,i}\,\varphi^i=0.\label{3.3}
\end{equation}
Here $\varphi^i$ now denotes the quantum field (variable of integration in the functional integral) and $\bar{K}_{\alpha\,i}$ is the gauge generator evaluated at the background field. There are two gauge conditions in our case that we will write in normal notation as
\begin{subequations}\label{3.4}
\begin{align}
C(x)&=\nabla^\mu A_\mu+i\,\zeta_1(\Phi^\dagger\bar{\Phi}-\bar{\Phi}^\dagger\Phi),\label{3.4a}\\
C_A(x)&=\nabla^\mu B_{A\mu}+i\,\zeta_2(\Phi^\dagger T_A\bar{\Phi}-\bar{\Phi}^\dagger T_A\Phi).\label{3.4b}
\end{align}
\end{subequations}
In order to show the gauge condition dependence of the usual formalism we have introduced two gauge parameters $\zeta_1$ and $\zeta_2$ here. The Landau-DeWitt gauge choice is specified by
\begin{subequations}\label{3.5}
\begin{align}
\zeta_1&=e_s,\label{3.5a}\\
\zeta_2&=g,\label{3.5b}
\end{align}
\end{subequations}
as follows directly from \eqref{3.3} by uncondensing the notation and using \eqref{2.7} and \eqref{3.2}. 

It is possible to keep $\zeta_1$ and $\zeta_2$ completely arbitrary. If this is done then it is necessary to allow for a non-trivial connection term in the Vilkovisky-DeWitt formalism above and beyond the Christoffel connection. This connection term enters the covariant differentiation in the expansion of the action about the background field and if done correctly completely cancels out the dependence on the parameters $\zeta_1$ and $\zeta_2$. The explicit demonstration of this was first given by Fradkin and Tseytlin~\cite{FradkinTseytlin} in the case of scalar electrodynamics. The demonstration of this in the present theory will be given elsewhere as a different approach to the calculations is more efficient for this. 

Because the formalism is completely independent of the gauge conditions, unlike the traditional methods, we can choose a particular gauge to simplify the calculations. The most convenient choice, as emphasised originally by Fradkin and Tseytlin~\cite{FradkinTseytlin}, is the Landau-DeWitt gauge condition specified by \eqref{3.4} and \eqref{3.5}. At one-loop order it is easy to see ({\em e.g.\/} \cite[p. 378]{ParkerTomsbook}) that the Landau-DeWitt gauge choice results in the extra terms in the Vilkovisky-DeWitt connection making no contribution to the result. For the present theory we also have the Christoffel term vanishing as mentioned above, so the complete connection makes no contribution. The net result is that we are able to adopt the conventional background field method provided that we adopt the Landau-DeWitt gauge conditions. Just to be clear, any choice of gauge conditions will lead to a result that coincides with that found by the traditional method in the Landau-DeWitt gauge; the Landau-DeWitt gauge choice is not mandatory for the calculation of the effective action but it makes the calculations far simpler. Our calculations will be in the Landau-DeWitt gauge \eqref{3.4} and \eqref{3.5} but we will relax the requirement of \eqref{3.5} until the end as this will show the gauge condition dependence of the traditional method.

The gauge conditions \eqref{3.4} can be incorporated into the functional integration that defines the effective action with Dirac $\delta$-functions along with the associated Faddeev-Popov determinant factors. The part of the one-loop effective action that arises from the scalar and vector fields, the Bose part $\Gamma^{(1)}_{\textrm{Bose}}$, is given by
\begin{equation}
e^{i\,\Gamma^{(1)}_{\textrm{Bose}}}=\int d\mu_{\textrm{Bose}}\,\delta[C(x)]\,\delta[C_A(x)]\,( \det\,{\mathcal{Q}})\,e^{i\,S_{\textrm{quad\,Bose}}},\label{3.6}
\end{equation}
where $d\mu_{\textrm{Bose}}$ indicates the functional integral measure over the Bose fields $\Phi,\Phi^\dagger,A_\mu,B_{A\mu}$, $\det\,{\mathcal{Q}}$ is the Faddeev-Popov determinant, and $S_{\textrm{quad\,Bose}}$ represents that part of the classical action that comes from the scalars and vectors and is quadratic in the quantum fields. With the gauge choice of \eqref{3.4} we have
\begin{equation}\label{3.7a}
\mathcal{Q}=\left( 
\begin{array}{cc}
Q_{x\,x'}&Q_{x\,Bx'}\\
Q_{Ax\,x'}&Q_{Ax\,Bx'}\\
\end{array}
\right),
\end{equation}
where
\begin{subequations}\label{3.7be}
\begin{align}
Q_{x\,x'}&= \Big(\Box_x+2\,\zeta_1 e_s\,|\bar{\Phi}|^2\Big)\delta(x,x'),\label{3.7b}\\
Q_{x\,Bx'}&=2\,\zeta_1 g\,\tau_B\,\delta(x,x'),\label{3.7c}\\
Q_{Ax\,x'}&= 2\,\zeta_2 e_s\,\tau_A\,\delta(x,x'),\label{3.7d}\\
Q_{Ax\,Bx'}&=\Big(\delta_{AB}\Box_x+2\,\zeta_2 g\,\rho_{AB}\Big)\delta(x,x').\label{3.7e}
\end{align}
\end{subequations}
We have defined
\begin{equation}
\tau_A=\bar{\Phi}^\dagger T_A\bar{\Phi},\label{3.8a}
\end{equation}
and
\begin{equation}
\rho_{AB}=\frac{1}{2}\,\bar{\Phi}^\dagger( T_AT_B+T_BT_A)\bar{\Phi},\label{3.8}
\end{equation}
as in \eqref{A10} and \eqref{A11}.

The spinor part of the one-loop effective action, $\Gamma^{(1)}_{\textrm{Fermi}}$, is given by
\begin{equation}
e^{i\,\Gamma^{(1)}_{\textrm{Fermi}}}=\int d\mu_{\textrm{Fermi}}\,e^{i\,S_{\textrm{quad\,spinor}}},\label{3.9}
\end{equation}
where $d\mu_{\textrm{Fermi}}$ represents the integration over the spinor fields $\chi,\bar{\chi},\Psi,\bar{\Psi}$, and $S_{\textrm{quad\,spinor}}$ is given by \eqref{2.4} with $\Phi=\bar{\Phi}$ taken there. The integration in \eqref{3.9} treats the fields as anti-commuting Grassmann variables. The result for $\Gamma^{(1)}_{\textrm{Fermi}}$ in \eqref{3.9} will be considered in Sec.~\ref{secspinors} after we have evaluated the contribution from Bose fields in the next section.

\section{Bose field contribution $\Gamma_{\textrm{Bose}}$}\label{secBose}

\subsection{Expansion of the effective action}\label{secexpansion}

From \eqref{3.6}, in order to evaluate the functional integral it is necessary to deal with the Dirac $\delta$-functions that encode the gauge conditions. The usual method is to exponentiate them using the functional equivalent of the identity
\begin{equation}
\delta(x)=\lim_{\alpha\rightarrow0}(-\pi i\alpha)^{-1/2}\,e^{-\,i\,\frac{x^2}{\alpha}}.\label{4.1}
\end{equation}
If this procedure is followed then the vector fields operators involve what was called non-minimal by Barvinsky and Vilkovisky~\citep{BarvinskyVilkovisky} where the second order derivative terms do not just involve $\Box=\nabla_\mu\nabla^\mu$ but can also have an explicit dependence on $\nabla_\mu\nabla_\nu$. Although it is perfectly possible to deal with such operators, there are extra complications in curved spacetime. (See for example, \cite{Tomsyukawa2} for how this works in a simple Abelian gauge theory.) What we will do instead avoids the need for non-minimal operators to a large degree, and also does not require the introduction of any extra gauge parameters like $\alpha$ in \eqref{4.1}.

In place of \eqref{4.1} we can use a Lagrange multiplier $\sigma$ and adopt the functional analogue of 
\begin{equation}
\delta(x)=\frac{1}{2\pi}\int d\sigma\,e^{i\sigma x}.\label{4.2}
\end{equation}
We can then take \eqref{3.6} to read
\begin{equation}
e^{i\,\Gamma^{(1)}_{\textrm{Bose}}}=( \det\,{\mathcal{Q}})\int d\tilde{\mu}_{\textrm{Bose}}\,e^{i\,\tilde{S}_{\textrm{quad\,Bose}}}.\label{4.3}
\end{equation}
Here 
\begin{equation}
\tilde{S}_{\textrm{quad\,Bose}}={S}_{\textrm{quad\,Bose}}+\int dv_x\Big\lbrack\sigma(x)C(x)+\sigma_A(x)C_A(x) \Big\rbrack,
\end{equation}
with $\sigma(x)$ and $\sigma_A(x)$ the Lagrange multiplier fields that enforce the gauge conditions \eqref{3.4}, and $d\tilde{\mu}_{\textrm{Bose}}$ indicates the functional integral measure over the Bose fields including the Lagrange multipliers, the full set being $\Phi,\Phi^\dagger,A_\mu,B_{A\mu},\sigma,\sigma_A$. It is advantageous to use the gauge conditions in \eqref{3.6} to simplify the quadratic part of the action by eliminating $(\nabla^\mu A_\mu)^2$ and $(\nabla^\mu B_{A\mu})^2$ in terms of $\Phi,\bar{\Phi}$ using \eqref{3.4}. This completely removes the non-minimal terms from the vector field operators, but of course there is a price to pay for this as we will see later.

It is convenient to write $\tilde{S}_{\textrm{quad\,Bose}}$ in the form
\begin{equation}
\tilde{S}_{\textrm{quad\,Bose}}=S_0+S_1+S_2,\label{4.5}
\end{equation}
where the subscripts $0,1,2$ indicate the power of the background scalar field $\bar{\Phi}$ that occurs. The explicit results for the three terms in \eqref{4.5} are
\begin{align}
S_0 &= \int dv_x \Big\lbrack\, \frac{1}{2}\,A^{\mu}\Box A_{\mu} + \frac{1}{2}\,R^{\mu\nu} A_{\mu} A_{\nu} +\frac{1}{2}\,B^{\mu}_{A}\Box B_{A\mu}+\frac{1}{2}\,R^{\mu\nu}B_{A\mu}B_{A\nu} \nonumber\\
&\qquad+\sigma_A\nabla^\mu B_{A\mu}+\sigma\nabla^\mu A_\mu+\nabla^\mu\Phi^\dagger\nabla_\mu\Phi-m_s^2\Phi^\dagger\Phi-\xi R \Phi^\dagger\Phi\Big\rbrack,\label{4.6}
\end{align}
that contains no dependence on the background scalar field,
\begin{align}
S_1&=\int dv_x\Big\lbrack 2\,ie_sA^\mu\Phi^\dagger\nabla_\mu\bar{\Phi} - 2\,ie_sA^\mu\nabla_\mu\Phi^\dagger\bar{\Phi}+2\,igB_{A\mu}\Phi^\dagger T_A\nabla_\mu\bar{\Phi} - 2\,igB_{A\mu}\nabla_\mu\Phi^\dagger T_A\bar{\Phi}\nonumber\\
&\qquad+i\zeta_1\sigma(\Phi^\dagger\bar{\Phi}-\bar{\Phi}^\dagger \Phi) +i\zeta_2\sigma_A(\Phi^\dagger T_A\bar{\Phi}-\bar{\Phi}^\dagger T_A\Phi) \Big\rbrack,\label{4.7}
\end{align}
that contains all terms with a linear dependence on $\bar{\Phi}$, and
\begin{align}
S_2&=\int dv_x\Big\lbrack \alpha_{ij}\,\Phi_i\Phi_j+\beta_{ij}\,\Phi_i^\dagger\Phi_j^\dagger+\gamma_{ij}\,\Phi_i^\dagger\Phi_j + e_s^2|\bar{\Phi}|^2\,A^\mu A_\mu\nonumber\\
&\qquad + 2\,e_sg\,\tau_A\,A^\mu B_{A\mu} + g^2\,\rho_{AB}\,B_{A}^{\mu}B_{A\mu}\Big\rbrack,\label{4.8}
\end{align}
that contains all terms quadratic in the background scalar field. We have used \eqref{3.8a} and \eqref{3.8} here and have defined
\begin{subequations}\label{4.9}
\begin{align}
\alpha_{ij}&=\Big(-\frac{1}{2}\zeta_1^2+\zeta_1e_s-\frac{\lambda}{6}\Big)\bar{\Phi}_i^\dagger \bar{\Phi}_j^\dagger + \Big(-\frac{1}{2}\zeta_2^2+\zeta_2 g\Big)(\bar{\Phi}^\dagger T_A)_i (\bar{\Phi}^\dagger T_A)_j\,\label{4.9a}\\
\beta_{ij}=(\alpha_{ij})^\dagger&=\Big(-\frac{1}{2}\zeta_1^2+\zeta_1e_s-\frac{\lambda}{6}\Big)\bar{\Phi}_i \bar{\Phi}_j + \Big(-\frac{1}{2}\zeta_2^2+\zeta_2 g\Big)(T_A\bar{\Phi} )_i (T_A\bar{\Phi})_j\,\label{4.9b}\\
\gamma_{ij}&=\Big(\zeta_1^2-2\,\zeta_1e_s-\frac{\lambda}{3}\Big)\bar{\Phi}_i \bar{\Phi}_j^\dagger + (\zeta_2^2-2\,\zeta_2 g)(T_A\bar{\Phi})_i (\bar{\Phi}^\dagger T_A)_j-\frac{\lambda}{3}\,\delta_{ij}\,|\bar{\Phi}|^2.\label{4.9c}
\end{align}
\end{subequations}
We can use \eqref{A9} to reduce these expressions slightly but we will not do this at this stage. We will wait until the end of the calculation and use \eqref{A10} and \eqref{A11} instead.

The aim now is to treat $S_1$ and $S_2$ as interactions and work to quartic order in $\bar{\Phi}$. The terms in $S_0$ determine the propagators to be used when the interaction terms are reduced using Wick's theorem. Only terms that correspond to connected one-particle irreducible diagrams need to be considered. Because $S_0$ involves terms that are not diagonal in the fields it is necessary to be careful here.

We can write
\begin{equation}
\Gamma^{(1)}_{\textrm{Bose}}=-i\ln\det{\mathcal Q}-i\left \langle e^{i(S_1+S_2)}\right\rangle,\label{4.10}
\end{equation}
where $\langle\cdots\rangle$ means to evaluate what is enclosed by the angle brackets using Wick's theorem with only one-particle irreducible terms kept. To quartic order in $\bar{\Phi}$ we have
\begin{equation}
\Gamma^{(1)}_{\textrm{Bose}}=-i\ln\det{\mathcal Q}+\Gamma_0+\Gamma_2+\Gamma_4+\cdots,\label{4.11}
\end{equation}
where $\Gamma_0$ denotes a vacuum term that is independent of $\bar{\Phi}$ that we will consider later, and
\begin{align}
\Gamma_2&=\langle S_2\rangle+\frac{i}{2}\langle S_1^2\rangle,\label{4.12}\\
\Gamma_4&=\frac{i}{2}\langle S_2\rangle-\frac{1}{2}\,\langle S_2 S_1^2\rangle - \frac{i}{24}\langle S_1^4\rangle,\label{4.13}
\end{align}
give the quadratic and quartic terms in $\bar{\Phi}$ coming from the scalars and vectors. There are also quadratic and quartic terms coming from the Faddeev-Popov ghost determinant that we will consider later in Sec.~\ref{secghost}. 

The Green's functions used in the evaluation of the terms in \eqref{4.12} and \eqref{4.13} follow from $S_0$ in \eqref{4.6}. The scalar fields occur without coupling to the other fields so we simply have
\begin{equation}
\langle\Phi_i(x)\Phi^\dagger_j(x')\rangle=i\,\delta_{ij}\,\Delta(x,x'),\label{4.14}
\end{equation}
where
\begin{equation}
(-\Box-m_s^2-\xi R)\,\Delta(x,x')=\delta(x,x').\label{4.15}
\end{equation}
(Clearly, $\langle\Phi_i(x)\Phi_j(x')\rangle=\langle\Phi^\dagger_i(x)\Phi^\dagger_j(x')\rangle=0$.) The vector fields and Lagrange multipliers are coupled together but with no coupling between the Abelian and non-Abelian parts. We will first consider the Abelian part.

Define the functional integral
\begin{equation}
Z\lbrack J^\mu,L\rbrack=\int\lbrack dA\,d\sigma\rbrack\,e^{iE},\label{4.16}
\end{equation}
where
\begin{equation}
E=\int dv_x\Big\lbrack\frac{1}{2}\,A^\mu(g_{\mu\nu}\Box+R_{\mu\nu})A^\nu+\sigma\nabla^\mu A_\mu +J^\mu A_\mu+L\sigma\Big\rbrack,\label{4.17}
\end{equation}
and $\lbrack dA\,d\sigma\rbrack$ represents the functional measure over the Abelian gauge field $A_\mu$ and the Lagrange multiplier field $\sigma$. Now define a Green's function $G^{\mu\nu}(x,x')$ by
\begin{equation}
(g_{\mu\nu}\Box+R_{\mu\nu})\,G^{\nu\lambda}(x,x')=\delta^{\lambda}_{\mu}\delta(x,x'),\label{4.18}
\end{equation}
so that $G^{\mu\nu}(x,x')$ may be viewed as the inverse of the operator $g_{\mu\nu}\Box+R_{\mu\nu}$. By operating on both sides of \eqref{4.18} with $\nabla^\mu\nabla^\prime_\lambda$ where $\nabla^\prime_\lambda$ is the covariant derivative at $x^\prime$, it is possible to show that
\begin{equation}
\nabla_\mu\nabla^\prime_\nu G^{\mu\nu}(x,x')=-\delta(x,x').\label{4.19}
\end{equation}
(This is trivial to show in flat spacetime; the curved spacetime result makes use of the contracted Bianchi identity.)

In \eqref{4.16} consider the integration over $A_\mu$ first and make the translation of integration variable
\begin{equation}
A_\mu(x)\rightarrow A_\mu(x)+\int dv_{x'}\,G^{\mu\nu}(x,x')\lbrack\nabla^\prime_\nu\sigma(x')-J_\nu(x')\rbrack,\label{4.20}
\end{equation}
which is chosen to eliminate the linear term in $A_\mu$ that occurs in \eqref{4.17}. After some manipulations involving integration by parts and use of \eqref{4.19} it can be shown that when \eqref{4.20} is made,
\begin{equation}
E\rightarrow\frac{1}{2}\int dv_x\,A^\mu(g_{\mu\nu}\Box+R_{\mu\nu})A^\nu - \frac{1}{2}\int dv_x dv_{x'}\,J^\mu(x)G_{\mu\nu}(x,x')J^\nu(x')+E_\sigma,\label{4.21}
\end{equation}
where $E_\sigma$ is given by
\begin{equation}
E_\sigma=\int dv_x\Big\lbrack\frac{1}{2}\,\sigma^2(x)+L(x)\sigma(x)\Big\rbrack - \int dv_x dv_{x'}\,\sigma(x)J_\mu(x')\nabla_\nu G^{\mu\nu}(x',x).\label{4.22}
\end{equation}
Now make the translation in the integration variable $\sigma(x)$ defined by
\begin{equation}
\sigma(x)\rightarrow \sigma(x)-L(x)+ \int dv_{x'}\,J_\mu(x')\nabla_\nu G^{\mu\nu}(x',x),\label{4.23}
\end{equation}
chosen to eliminate the term linear in $\sigma(x)$ in \eqref{4.22}. After a bit of calculation it follows that \eqref{4.16} becomes
\begin{equation}
Z\lbrack J^\mu,L\rbrack=Z\lbrack 0,0\rbrack\,e^{i{\mathcal E}},\label{4.24}
\end{equation}
where
\begin{align}
{\mathcal E}&=- \int dv_x dv_{x'}\,J_\mu(x) \widetilde{G}^{\mu\nu}(x,x')J_\nu(x')\nonumber\\
&\qquad + 2\,\int dv_x dv_{x'}\,J_\mu(x')\nabla_\nu {G}^{\mu\nu}(x',x)L(x)-\int dv_x\,L^2(x).\label{4.25}
\end{align}
($Z\lbrack0,0\rbrack$ will be evaluated in Sec.~\ref{secvacuum} as it is not needed here.) We have defined $\widetilde{G}^{\mu\nu}(x,x')$ by
\begin{equation}
\widetilde{G}^{\mu\nu}(x,x')={G}^{\mu\nu}(x,x')+\int dv_{x''}\nabla^{\prime\prime}_\lambda G^{\mu\lambda}(x,x'')\nabla^{\prime\prime}_\sigma G^{\nu\sigma}(x',x'').\label{4.26}
\end{equation}

The two-point functions are now easily evaluated from \eqref{4.24} and \eqref{4.25}. For example,
\begin{align}
\langle A_\mu(x) A_\nu(x')\rangle &=\frac{1}{Z\lbrack0,0\rbrack}\int\lbrack dA\, d\sigma\rbrack\,A_\mu(x)A_\nu(x')e^{iS_0}\nonumber\\
&=\left.\frac{1}{Z\lbrack0,0\rbrack}\left(-i\frac{\delta}{\delta J^\mu(x)}\right)\left(-i\frac{\delta}{\delta J^\nu(x')}\right)Z\lbrack J^\mu,L\rbrack\right|_{J^\mu=L=0}\nonumber\\
&= i\,\widetilde{G}^{\mu\nu}(x,x').\label{4.27}
\end{align}
In a similar way,
\begin{equation}
\langle A_\mu(x) \sigma(x')\rangle=-i\,G_\mu(x,x'),\label{4.28}
\end{equation}
where
\begin{equation}
G_\mu(x,x')=\nabla^{\prime\nu}G_{\mu\nu}(x,x').\label{4.29}
\end{equation}
Finally,
\begin{equation}
\langle \sigma(x) \sigma(x')\rangle=i\,\delta(x,x').\label{4.30}
\end{equation}

A similar analysis to that just described applied to the Yang-Mills field gives
\begin{subequations}\label{4.31}
\begin{align}
\langle B_{A\mu}(x) B_{B\nu}(x')\rangle &= i\,\delta_{AB}\,\widetilde{G}^{\mu\nu}(x,x'),\label{4.31a}\\
\langle B_{A\mu}(x) \sigma_B(x')\rangle&=-i\,\delta_{AB}\,G_\mu(x,x'),\label{4.31b}\\
\langle \sigma_A(x) \sigma_B(x')\rangle&=i\,\delta_{AB}\,\delta(x,x').\label{4.31c}
\end{align}
\end{subequations}

The results in \eqref{4.14},\eqref{4.27},\eqref{4.28},\eqref{4.30} and \eqref{4.31} enable all of the terms in the one-loop effective action to be determined. Before proceeding with the evaluation of the one-loop effective action it is advantageous to put \eqref{4.26} into a more convenient form. This is described in Appendix~\ref{appGreen}.

\subsection{Poles in $\Gamma_2$}\label{secquad}

It is now a straightforward matter to evaluate the pole part of $\Gamma_2$ defined in \eqref{4.12}. From \eqref{4.8} we have
\begin{align}
\langle S_2\rangle&=\int dv_x\Big\lbrack\gamma_{ij}\langle \Phi^\dagger_i\Phi_j\rangle +e_s^2|\bar{\Phi}|^2\langle A_\mu A_\nu\rangle +g^2\rho_{AB}\langle B_A^\mu B_{B\mu}\rangle\Big\rbrack\nonumber\\
&=i\,\int dv_x\Big\lbrack\gamma_{ii}\Delta(x,x) +\Big(e_s^2|\bar{\Phi}|^2+g^2\rho_{AA}\Big)\,\widetilde{G}^{\mu}{}_{\mu}(x,x) \Big\rbrack,\label{4.44}
\end{align}
where \eqref{4.14}, \eqref{4.27} and \eqref{4.31a} were used.

If ${\textrm{PP}}\lbrace\cdots\rbrace$ denotes the pole part of any expression in dimensional regularisation then using \eqref{B2} and \eqref{B3} it can be shown that
\begin{align}
{\textrm{PP}}\lbrace\langle S_2\rangle\rbrace &=\frac{1}{8\pi^2\epsilon}\int dv_x\Big\lbrace\Big\lbrack \zeta_1^2-2\,\zeta_1e_s+\zeta_2^2\,C_2-2\,\zeta_2\,g\,C_2-\frac{\lambda}{3}(d_R+1)\Big\rbrack\,m_s^2\,|\bar{\Phi}|^2\label{4.45}\\
&\Big\lbrack \zeta_1^2-2\,\zeta_1e_s+\zeta_2^2\,C_2-2\,\zeta_2\,g\,C_2-\frac{\lambda}{3}(d_R+1)\Big\rbrack\left(\xi-\frac{1}{6}\right)-\frac{1}{2}\,e_s^2-\frac{1}{2}\,g^2\,C_2\Big\rbrack\,R\,|\bar{\Phi}|^2\Big\rbrace.\nonumber
\end{align}
We have just written $C_2$ in place of $C_2(G_R)$ since no confusion should arise as we only consider one representation here. The results of \eqref{4.9c} and \eqref{3.8} have been used along with \eqref{A4}.

Turning next to $\langle S_1^2\rangle$ we use \eqref{4.7} and the result is somewhat more complicated than that for $\langle S_2\rangle$. For convenience it proves useful to write $S_1$ in the form
\begin{equation}
S_1=S_{11}+S_{12}+\cdots +S_{16},\label{4.46}
\end{equation}
where $S_{11},\ldots,S_{16}$ represent the six terms that occur on the right hand side of \eqref{4.7}. A consideration of the non-trivial pairings in $\langle S_1^2\rangle$ shows that
\begin{align}
\langle S_1^2\rangle&=2\,\langle S_{11}S_{12}\rangle + 2\,\langle S_{11}S_{15}\rangle + 2\,\langle S_{12}S_{15}\rangle+ 2\,\langle S_{13}S_{14}\rangle+2\,\langle S_{13}S_{16}\rangle+2\,\langle S_{14}S_{16}\rangle\nonumber\\
&\quad+ \langle S_{15}^2\rangle + \langle S_{16}^2\rangle.\label{4.47}
\end{align}
From the first two terms on the right hand side of \eqref{4.7} it can be shown that
\begin{equation}
\langle S_{11}S_{12}\rangle=-4\,e_s^2\,\int dv_xdv_{x'}\nabla^\prime_\nu\bar{\Phi}^\dagger(x')\nabla_\mu\bar{\Phi}(x)\widetilde{G}^{\mu\nu}(x,x')\Delta(x,x'),\label{4.48}
\end{equation}
if we use \eqref{4.14} and \eqref{4.27}. Using \eqref{B4} shows that
\begin{equation}
{\textrm{PP}}\lbrace \langle S_{11}S_{12}\rangle\rbrace=-\,\frac{3\,i\,e_s^2}{8\pi^2\epsilon}\,\int dv_x\nabla^\mu\bar{\Phi}^\dagger\nabla_\mu\bar{\Phi}.\label{4.49}
\end{equation}

From the first and fifth terms in \eqref{4.7} it follows that
\begin{displaymath}
\langle S_{11}S_{15}\rangle=2\,\zeta_1\,e_s\,\int dv_xdv_{x'}\bar{\Phi}^\dagger(x')\nabla_\mu\bar{\Phi}(x){G}^{\mu}(x,x')\Delta(x,x'),
\end{displaymath}
if we use \eqref{4.14} and \eqref{4.28}. From \eqref{B5} we find
\begin{equation}
{\textrm{PP}}\lbrace \langle S_{11}S_{15}\rangle\rbrace=-\,\frac{i\,\zeta_1\,e_s}{8\pi^2\epsilon}\,\int dv_x\nabla^\mu\bar{\Phi}^\dagger\nabla_\mu\bar{\Phi}.\label{4.50}
\end{equation}
In a similar way, and as is obvious from symmetry,
\begin{equation}
{\textrm{PP}}\lbrace \langle S_{12}S_{15}\rangle\rbrace=-\,\frac{i\,\zeta_1\,e_s}{8\pi^2\epsilon}\,\int dv_x\nabla^\mu\bar{\Phi}^\dagger\nabla_\mu\bar{\Phi}.\label{4.51}
\end{equation}

For $\langle S_{13}S_{14}\rangle$ we need \eqref{4.14} and \eqref{4.31a}. The definition in \eqref{A4} is also required. It is found that
\begin{equation}
{\textrm{PP}}\lbrace \langle S_{13}S_{14}\rangle\rbrace=-\,\frac{3\,i\,g^2\,C_2}{8\pi^2\epsilon}\,\int dv_x\nabla^\mu\bar{\Phi}^\dagger\nabla_\mu\bar{\Phi}.\label{4.52}
\end{equation}

For  $\langle S_{13}S_{15}\rangle$ we use \eqref{4.14} and \eqref{4.31b} along with \eqref{A4} to obtain
\begin{equation}
{\textrm{PP}}\lbrace \langle S_{13}S_{15}\rangle\rbrace=-\,\frac{i\,\zeta_2\,g\,C_2}{8\pi^2\epsilon}\,\int dv_x\nabla^\mu\bar{\Phi}^\dagger\nabla_\mu\bar{\Phi}.\label{4.53}
\end{equation}
In a similar way, or by symmetry,
\begin{equation}
{\textrm{PP}}\lbrace \langle S_{14}S_{16}\rangle\rbrace=-\,\frac{i\,\zeta_2\,g\,C_2}{8\pi^2\epsilon}\,\int dv_x\nabla^\mu\bar{\Phi}^\dagger\nabla_\mu\bar{\Phi}.\label{4.54}
\end{equation}

For $\langle S_{15}^2\rangle$ we need to use \eqref{4.14} and \eqref{4.30}. It is found that
\begin{displaymath}
 \langle S_{15}^2\rangle=-2\,\zeta_1^2\,\int dv_x |\bar{\Phi}|^2\Delta(x,x).
\end{displaymath}
Use of \eqref{B2} shows that
\begin{equation}
{\textrm{PP}}\lbrace \langle S_{15}^2\rangle\rbrace=\frac{i\zeta_1^2}{4\pi^2\epsilon}\,\int dv_x \Big\lbrack m_s^2\,|\bar{\Phi}|^2+\Big(\xi-\frac{1}{6}\Big)R\,|\bar{\Phi}|^2\Big\rbrack.\label{4.56}
\end{equation}
A similar result to this holds for $\langle S_{16}^2\rangle$ with an extra factor of the Casimir invariant $C_2$ coming from the use of \eqref{A4}:
\begin{equation}
{\textrm{PP}}\lbrace \langle S_{16}^2\rangle\rbrace=\frac{i\zeta_2^2\,C_2}{4\pi^2\epsilon}\,\int dv_x \Big\lbrack m_s^2\,|\bar{\Phi}|^2+\Big(\xi-\frac{1}{6}\Big)R\,|\bar{\Phi}|^2\Big\rbrack.\label{4.57}
\end{equation}

Using \eqref{4.49}--\eqref{4.57} back in \eqref{4.47} shows that
\begin{align}
{\textrm{PP}}\lbrace \langle S_{1}^2\rangle\rbrace&= \frac{i}{8\pi^2\epsilon}\,\int dv_x\Big\lbrace -(6\,e_s^2+4\,\zeta_1\,e_s+6\,g^2\,C_2+4\,\zeta_2\,g\,C_2)\nabla^\mu\bar{\Phi}^\dagger\nabla_\mu\bar{\Phi} 
\nonumber\\
&\quad+(2\,\zeta_1^2+2\,\zeta_2^2\,C_2)\Big\lbrack m_s^2+\Big(\xi-\frac{1}{6}\Big)R\Big\rbrack\,|\bar{\Phi}|^2\Big\rbrace.\label{4.58}
\end{align}

If we now combine \eqref{4.45} with \eqref{4.58} we will obtain the pole part of $\Gamma_2$ from \eqref{4.12} to be
\begin{align}
{\textrm{PP}}\lbrace \Gamma_2\rbrace&= \frac{1}{8\pi^2\epsilon}\,\int dv_x\Big\lbrace (3\,e_s^2+2\,\zeta_1\,e_s+3\,g^2\,C_2+2\,\zeta_2\,g\,C_2)\nabla^\mu\bar{\Phi}^\dagger\nabla_\mu\bar{\Phi} 
\nonumber\\
&\qquad-\Big\lbrack 2\,\zeta_1\,e_s+2\,\zeta_2\,g\,\,C_2+\frac{\lambda}{3}(d_R+1)\Big\rbrack\Big\lbrack m_s^2+\Big(\xi-\frac{1}{6}\Big)R\Big\rbrack\,|\bar{\Phi}|^2\nonumber\\
&\qquad - \frac{1}{2}(e_s^2+g^2\,C_2)R \,|\bar{\Phi}|^2\Big\rbrace.\label{4.59}
\end{align}
There are no other sources for pole terms that are quadratic in the background scalar fields in the traditional effective action method, so this clearly shows the gauge condition dependence of the result through the dependence on the two parameters $\zeta_1$ and $\zeta_2$. As mentioned, in the Vilkovisky-DeWitt method there will be extra contributions coming from the additional terms in the connection that will cancel these gauge condition parameters to leave a result that coincides with the Landau-DeWitt gauge condition given in \eqref{3.5} and which is free of any dependence on these parameters.

\subsection{Poles in $\Gamma_4$}\label{secquartic}

Some details in the evaluation of the three terms in \eqref{4.13} are contained in Appendix~\ref{appGamma4}. If we use \eqref{C14}, \eqref{C16} and \eqref{C18} then we can write
\begin{equation}
{\textrm{PP}}\lbrace \Gamma_4\rbrace=\frac{1}{8\pi^2\epsilon}\,\int dv_x\Big( A\,|\bar{\Phi}|^4+B\,\tau_A^2+D\,\rho_{AB}^2\Big),\label{4.60}
\end{equation}
with $\tau_A$ and $\rho_{AB}$ given by \eqref{3.8a} and \eqref{3.8}. The three coefficients in \eqref{4.60} are found to be
\begin{subequations}\label{4.61}
\begin{align}
A&=-3\,e_s^4-2\,\zeta_1^2\,e_s^2-\frac{2\,\lambda}{3}\,\left( \zeta_1\,e_s+C_2\,\zeta_2\,g\right) - \frac{(d_R+4)}{18}\,\lambda^2,\label{4.61a}\\
B&=-6\,e_s^2\,g^2-4\,\zeta_1\,\zeta_2\,e_s\,g,\label{4.61b}\\
D&=-3\,g^4-2\,\zeta_2^2\,g^2.\label{4.61c}
\end{align}
\end{subequations}

As a check on these results we can compare them with the Abelian case in the Landau-DeWitt gauge calculated in \cite{Tomsyukawa2}. If we set $\zeta_2=g=0,\ d_R=1$ and $\zeta_1=e_s$ then $A=-5\,e_s^4-2\,\lambda\,e_s^2/3-5\,\lambda^2/18,\ B=D=0$ which does agree with the analogous result of \cite{Tomsyukawa2}. We can also check agreement by taking $\zeta_1=e_s=0,\ \zeta_2=g$ and $d_R=1$ along with $T_A\rightarrow1$. In this case we have $A=-2\,\lambda\,g^2/3-5\,\lambda^2/18,\ B=0,\ D=-5\,g^4$ so that the overall coefficient of $|\bar{\Phi}|^4$ is $A+D=-5\,g^4-2\,\lambda\,g^2/3-5\,\lambda^2/18$ just as before.

The overall coefficient of $|\bar{\Phi}|^4$ in \eqref{4.60} is found by using \eqref{A10} and \eqref{A11} to reduce the second and third terms. However before doing this we will consider the ghost contribution $-i\,\ln\det{\mathcal Q}$ coming from the first term of \eqref{4.11}.

\subsection{Ghost contribution}\label{secghost}

The ghost contribution to the one-loop effective action is given by
\begin{equation}
\Gamma_{\textrm{ghost}}=-i\,\ln\det{\mathcal Q},\label{4.4.1}
\end{equation}
where ${\mathcal Q}$ was defined in \eqref{3.7a} with \eqref{3.7be}. It is most expedient to use heat kernel methods here since this will also give us the vacuum contributions as well that depend on the spacetime curvature. We will indicate how the pole parts of \eqref{4.4.1} that involve quadratic and quartic terms in $\bar{\Phi}$ may also be found perturbatively.

The operator ${\mathcal Q}$ was defined in \eqref{3.7a} takes the general form
\begin{equation}
{\mathcal Q}=\left(\begin{array}{cc}\Box_x&0\\ 0&\delta_{AB}\,\Box_x\\ \end{array}\right)\delta(x,x') + \left(\begin{array}{cc}X_{x\,x}&X_{x\,Bx}\\ X_{Ax\,x}&X_{Ax\,Bx}\\ \end{array}\right)\delta(x,x'),\label{4.4.2}
\end{equation}
where
\begin{subequations}\label{4.4.3}
\begin{align}
X_{x\,x}&=2\,\zeta_1\,e_s|\bar{\Phi}|^2,\label{4.4.3a}\\
X_{x\,Bx}&=2\,\zeta_1\,g\,\tau_{B},\label{4.4.3b}\\
X_{Ax\,x}&=2\,\zeta_2\,e_s\,\tau_{A},\label{4.4.3c}\\
X_{Ax\,Bx}&=2\,\zeta_2\,g\,\rho_{AB}.\label{4.4.3d}
\end{align}
\end{subequations}
For any covariant derivative $D_\mu$ and any $X(x)$ we have (using the notation of \cite[pages 193--194]{ParkerTomsbook})
\begin{equation}
\textrm{PP}\left\lbrace i\,\ln\det(D^2+X)\right \rbrace=-\frac{1}{8\pi^2\epsilon}\int dv_x\,\textrm{tr}\,E_2(x),\label{4.4.4}
\end{equation}
where
\begin{align}
E_2&=\left(\frac{1}{72}\,R^2-\frac{1}{180}\,R^{\mu\nu}R_{\mu\nu}+\frac{1}{180}\,R^{\mu\nu\lambda\sigma}R_{\mu\nu\lambda\sigma}\right)\,I\nonumber\\
&\quad +\frac{1}{12}\,W^{\mu\nu}W_{\mu\nu}+\frac{1}{2}\,X^2-\frac{1}{6}\,R\,X.\label{4.4.5}
\end{align}
Here $W_{\mu\nu}=\lbrack D_\mu,D_\nu\rbrack=0$ since the ghost fields are scalars, but we quote the more general result for later reference. A total derivative of $X$ in \eqref{4.4.5} has been discarded as it cannot contribute to \eqref{4.4.4}. (See \cite{DeWittdynamical,Gilkey75,Gilkey79} for original calculations of the heat kernel coefficients or \cite{fulling1989aspects,avramidi2000heat,vassilevich2003heat,ParkerTomsbook,kirsten2010spectral} for reviews.) For the ghost operator in \eqref{4.4.2} we have
\begin{align}
{\textrm {tr}}\,I&=1+d_R,\label{4.4.6}\\
{\textrm {tr}}\,X&=2\,\zeta_1\,e_s\,|\bar{\Phi}|^2+2\,\zeta_2\,g\,C_2\,|\bar{\Phi}|^2, 
\label{4.4.7}\\
{\textrm {tr}}\,X^2&=4\,\zeta_1^2\,e_s^2\,|\bar{\Phi}|^4+ 8\,\zeta_1\,\zeta_2\,e_s\,g\,\tau_A^2
+4\,\zeta_2^2\,g^2\,\rho_{AB}^2.\label{4.4.8}
\end{align}

We then find
\begin{align}
{\textrm{PP}}\lbrace \Gamma_{\textrm{ghost}}\rbrace&=\frac{1}{8\pi^2\epsilon}\,\int dv_x\Big\lbrack (d_R+1)\Big( \frac{1}{72}\,R^2-\frac{1}{180}\,R^{\mu\nu}R_{\mu\nu}+ \frac{1}{180}\,R^{\mu\nu\lambda\sigma}R_{\mu\nu\lambda\sigma}\Big)\nonumber\\
&\qquad\qquad-\frac{1}{3}\,(\zeta_1\,e_s+C_2\,\zeta_2\,g)\,R\,|\bar{\Phi}|^2 + 2\,\zeta_1^2\,e_s^2\,|\bar{\Phi}|^4\nonumber\\
&\qquad\qquad+4\,\zeta_1\,\zeta_2\,e_s\,g\,\tau_A^2+2\,\zeta_2^2\,g^2\,\rho_{AB}^2\Big\rbrack.\label{4.4.9}
\end{align}
As with previous results, the final three terms in \eqref{4.4.9} can be combined into a single term that involves $|\bar{\Phi}|^4$ if we use \eqref{A10} and \eqref{A11}.

The last four terms of \eqref{4.4.9} can also be evaluated by more traditional methods along the lines of our previous approach. Define anticommuting Faddeev-Popov ghosts $\theta(x)$ and $\theta_A(x)$. Then we can write \eqref{4.4.1} as
\begin{equation}
e^{i\,\Gamma_{\textrm{ghost}}}=\int\lbrack d\theta\,d\theta_A\rbrack\,e^{i\,S_{\textrm{ghost}}},\label{4.4.10}
\end{equation}
where
\begin{subequations}\label{4.4.11}
\begin{align}
S_{\textrm{ghost}}&=\int dv_x\Big(\bar{\theta}\,\Box\,\theta+\bar{\theta}_A\,\Box\,\theta_A\Big)+ S_{\textrm{ghost}}^{\textrm{int}},\label{4.4.11a}\\
S_{\textrm{ghost}}^{\textrm{int}}&=\int dv_x\Big(2\,\zeta_1\,e_s\,|\bar{\Phi}|^2\,\bar{\theta}\,\theta + 2\,\zeta_1\,g\,\tau_A\,\bar{\theta}\,\theta_A\nonumber\\
&\qquad + 2\,\zeta_2\,e_s\,\tau_A\,\bar{\theta}_A\,\theta + 2\,\zeta_2\,g\,\rho_{AB}\,\bar{\theta}_A\,\theta_B\Big).\label{4.4.11b}
\end{align}
\end{subequations}
$S_{\textrm{ghost}}$ is treated as the action for the ghost fields, with \eqref{4.4.11b} treated as an interaction. It then follows that, with the neglect of the vacuum terms,
\begin{equation}
{\textrm{PP}}\lbrace \Gamma_{\textrm{ghost}}\rbrace={\textrm{PP}}\left\lbrace \left\langle S_{\textrm{ghost}}^{\textrm{int}}\right\rangle+ \frac{i}{2}\, \left\langle \left(S_{\textrm{ghost}}^{\textrm{int}}\right)^2\right\rangle\right\rbrace.\label{4.4.12}
\end{equation}
The basic relations needed here are
\begin{subequations}\label{4.4.13}
\begin{align}
\langle\theta(x)\,\bar{\theta}(x')\rangle&=-i\,\Delta_{\textrm{g}}(x,x'),\label{4.4.13a}\\
\langle\theta_A(x)\,\bar{\theta}_B(x')\rangle&=-i\,\delta_{AB}\,\Delta_{\textrm{g}}(x,x'),\label{4.4.13b}
\end{align}
\end{subequations}
where
\begin{equation}
-\Box\,\Delta_{\textrm{g}}(x,x')=\delta(x,x').\label{4.4.13c}
\end{equation}
The local momentum space expansion for $\Delta_{\textrm{g}}(x,x')$ follows from taking $m_s^2=0$ and $\xi=0$ in the Green's function for the scalar field given in \eqref{4.43}. In particular we find
\begin{align}
{\textrm{PP}}\lbrace \Delta_{\textrm{g}}(x,x)\rbrace&=\frac{i}{48\pi^2\epsilon}\,R,\label{4.4.14a}\\
{\textrm{PP}}\lbrace \Delta_{\textrm{g}}^2(x,x')\rbrace&=-\,\frac{i}{8\pi^2\epsilon}\,\delta(x,x'),\label{4.4.14b}
\end{align}
from \eqref{B2} and \eqref{B6}.

It is now straightforward to show that
\begin{align}
\left\langle S_{\textrm{ghost}}^{\textrm{int}}\right\rangle&=2i\,
\int dv_x\left(\zeta_1\,e_s+C_2\,\zeta_2\,g\right)\,|\bar{\Phi}|^2\,\Delta_{\textrm{g}}(x,x),\label{4.4.15a}\\
\left\langle \left(S_{\textrm{ghost}}^{\textrm{int}}\right)^2\right\rangle&=\int dv_x dv_{x'}\Big\lbrack 4\,\zeta_1^2\,e_s^2\,|\bar{\Phi}(x)|^2\,|\bar{\Phi}(x')|^2 +8\,\zeta_1\,\zeta_2\,e_s\,g\,\tau_A(x)\,\tau_A(x')\nonumber\\
&\qquad + 4\,\zeta_2^2\,g^2\,\rho_{AB}(x)\,\rho_{AB}(x')\Big\rbrack\,\Delta_{\textrm{g}}^2(x,x').\label{4.4.15b}
\end{align}
Upon use of \eqref{4.4.14a} and \eqref{4.4.14b} in these last two results, and substitution back into \eqref{4.4.12}, the last four terms of \eqref{4.4.9} are found as claimed.

\subsection{Vacuum contribution}\label{secvacuum}

To complete the pole part of the one-loop effective action we need the vacuum part that is independent of the background scalar fields and depends only on the curvature of the background spacetime. This follows directly by performing the integration over the Bose fields using \eqref{4.6} and adding in the ghost contribution from \eqref{4.4.9}. We can define the vacuum Bose field contribution to be $\Gamma_{\textrm{Bose}}^{\textrm{vac}}$ where
\begin{equation}
e^{i\,\Gamma_{\textrm{Bose}}^{\textrm{vac}}}=\int d\mu\,e^{i\,S_0}.\label{4.5.1}
\end{equation}
Here $d\mu$ stands for the functional measure over the Bose fields $\Phi^\dagger,\Phi,A_\mu,B_{A\mu},\sigma$ and $\sigma_A$. The form of $S_0$ in \eqref{4.6} shows that the functional integral factorises into three: one over $\Phi^\dagger$ and $\Phi$, one over $A_\mu$ and $\sigma$, and one over $B_{A\mu}$ and $\sigma_A$. The integral over $\Phi^\dagger$ and $\Phi$ gives rise to $\lbrack\det(\Box+m_s^2+\xi R)\rbrack^{-1}$. The integral over $A_\mu$ and $\sigma$ can be done as described in Sec.~\ref{secexpansion} and coincides with $Z\lbrack0,0\rbrack$ defined in \eqref{4.16} with \eqref{4.17}. Making the change of variables in \eqref{4.20} with $J_\nu=0$ taken there results in 
\begin{equation}
Z\lbrack0,0\rbrack=\Big\lbrack \det\Big(\delta^\mu_\nu\,\Box+R^{\mu}_{\nu}\Big)\Big\rbrack^{-1/2}.\label{4.5.2}
\end{equation}
up to an overall constant. The integral over $B_{A\mu}$ and $\sigma_A$ may be done in the same way and gives $\lbrack\det(\delta^\mu_\nu\,\Box+R^{\mu}_{\nu})\rbrack^{-d_R/2}$ allowing for the extra gauge indices. The net result is that
\begin{equation}
\Gamma_{\textrm{Bose}}^{\textrm{vac}}=i\,\ln\det\Big(\Box+m_s^2+\xi R\Big)+ \frac{i}{2}\,(d_R+1)\, \ln\det\Big(\delta^\mu_\nu\,\Box+R^{\mu}_{\nu}\Big).\label{4.5.3}
\end{equation}
We now use the basic heat kernel result \eqref{4.4.4} with \eqref{4.4.5}. For the scalar fields $W_{\mu\nu}=0$ and $X_{ij}=(m_s^2+\xi R)\,\delta_{ij}$. For the vector fields $(W_{\mu\nu})^{\lambda}{}_{\sigma}=-R_{\mu\nu}{}^{\lambda}{}_{\sigma}$ and $X^{\mu}{}_{\nu}=R^{\mu}_{\nu}$. The net result is that
\begin{align}
{\textrm{PP}}\lbrace \Gamma_{\textrm{Bose}}^{\textrm{vac}}\rbrace&=-\frac{1}{8\pi^2\epsilon}\,\int dv_x\Big\lbrace \frac{1}{18}\,\Big\lbrack 9\,d_R\,\Big(\xi-\frac{1}{6}\Big)^2-d_R-1\Big\rbrack\,R^2\nonumber\\
&+\frac{1}{180}\,(42\,d_R+43)\,R^{\mu\nu}R_{\mu\nu}-\frac{1}{360}\,(9\,d_R+11)\,R^{\mu\nu\lambda\sigma}R_{\mu\nu\lambda\sigma}\nonumber\\
&+\frac{1}{2}\,d_R\,m_s^4+d_R\,\Big(\xi-\frac{1}{6}\Big)\,m_s^2\,R\Big\rbrace.\label{4.5.4}
\end{align}
is the expression for the pole part of the one-loop effective action that involves just the Bose fields and contains no terms in the background scalar field.

\subsection{Complete expression for the one-loop scalar pole terms}\label{secBosecomplete}

In this section we will combine all of the separate pieces of the pole terms for the one-loop effective action worked out above. Using \eqref{4.59}, \eqref{4.60}, \eqref{4.4.9} and \eqref{4.5.4} along with \eqref{A10} and \eqref{A11} the following result is obtained:
\begin{align}
{\textrm{PP}}\lbrace \Gamma_{\textrm{Bose}}^{(1)}\rbrace&=\frac{1}{8\pi^2\epsilon}\,\int dv_x\Big\lbrace(3\,e_s^2+2\,\zeta_1\,e_s+3\,g^2\,C_2+2\,\zeta_2\,g\,C_2)\nabla^\mu\bar{\Phi}^\dagger\nabla_\mu\bar{\Phi} 
\nonumber\\
&\qquad-\Big\lbrack 2\,\zeta_1\,e_s+2\,\zeta_2\,g\,\,C_2+\frac{\lambda}{3}(d_R+1)\Big\rbrack\, m_s^2\,|\bar{\Phi}|^2\nonumber\\
&\qquad-\Big\lbrack \frac{1}{2}\,e_s^2+\frac{1}{2}\,g^2\,C_2+ \frac{1}{3}\,\zeta_1\,e_s+\frac{1}{3}\,C_2\,\zeta_2\,g\nonumber\\
&\qquad\qquad+  \Big(\xi-\frac{1}{6}\Big)\Big( 2\,\zeta_1\,e_s+2\,\zeta_2\,g\,\,C_2+\frac{\lambda}{3}(d_R+1)\Big)\Big\rbrack\,R\,|\bar{\Phi}|^2\nonumber\\
&\qquad-\Big\lbrack 3\,e_s^4+\frac{2\,\lambda}{3}\,\left( \zeta_1\,e_s+C_2\,\zeta_2\,g\right) + \frac{(d_R+4)}{18}\,\lambda^2\nonumber\\
&\qquad+6\,C_2\,e_s^2\,g^2\left(\frac{d_R-1}{d_G}\right) +3\,C_2^2\,g^4 
\left(\frac{(d_R-1)(d_R^2+2\,d_R-2)}{2\,d_G^2}\right)  \Big\rbrack\,|\bar{\Phi}|^4\nonumber\\
&\qquad+\frac{1}{72}\,\Big\lbrack 5\,d_R+5 -36\,d_R\,\Big(\xi-\frac{1}{6}\Big)^2\Big\rbrack\,R^2\nonumber\\
&\qquad-\frac{1}{180}\,(43\,d_R+44)\,R^{\mu\nu}R_{\mu\nu}+\frac{1}{360}\,(11\,d_R+13)\,R^{\mu\nu\lambda\sigma}R_{\mu\nu\lambda\sigma}\nonumber\\
&\qquad-\frac{1}{2}\,d_R\,m_s^4-d_R\,\Big(\xi-\frac{1}{6}\Big)\,m_s^2\,R\Big\rbrace.\label{4.6.1}
\end{align} 
Some, but not all, of the dependence on the gauge parameters $\zeta_1$ and $\zeta_2$ has cancelled out here. In particular, $\zeta_1$ and $\zeta_2$ enter the term that involves the derivatives of the background scalar field and will therefore enter into the field renormalisation in the conventional method. Of course in our case we have the Landau-DeWitt gauge $\zeta_1=e_s$ and $\zeta_2=g$ that will coincide with the gauge condition independent result of the Vilkovisky-DeWitt method. If we specialise to the gauge group $SU(n)$, then the term in $g^4$ that multiplies $|\bar{\Phi|^4}$ agrees precisely with that found in \cite{Chengetal}. This provides a useful check on our calculations.

\section{Spinor field contribution $\Gamma^{(1)}_{\textrm{Fermi}}$}\label{secspinors}

Referring back to \eqref{3.9} we can perform the integration over the Grassmann variables $\Psi$ and $\chi$ to obtain the the contribution of fermions to the one-loop effective action as the functional determinant
\begin{equation}
\Gamma_{\textrm{fermion}}=-i\,\ln\det\,\left( \begin{array}{cc} \mathbb{A}(x,x')&\mathbb{B}(x,x')\\ \mathbb{C}(x,x')&\mathbb{D}(x,x')\end{array}\right),\label{F1}
\end{equation}
where
\begin{subequations}\label{F2}
\begin{align}
\mathbb{A}(x,x')&=\Big(i\gamma^\mu \nabla_{\!\!\mu}-m_{\psi}-im_{\psi5}\gamma_5\Big)\delta(x,x'),\label{F2a}\\
\mathbb{B}(x,x')&=-\,\bar{\Phi}(x)\,(w^\ast+iw_5^\ast\gamma_5)\,\delta(x,x'),\label{F2b}\\
\mathbb{C}(x,x')&=-\,(w+iw_5\gamma_5)\,\bar{\Phi}^\dagger(x)\delta(x,x'),\label{F2c}\\
\mathbb{D}(x,x')&=\Big(i\gamma^\mu \nabla_{\!\!\mu}-m_{\chi}-im_{\chi5}\gamma_5\Big)\delta(x,x').\label{F2d}
\end{align}
\end{subequations}
The results in \eqref{F2} follow directly from the spinor field action given in \eqref{2.4} with the scalar field $\Phi$ taken to be the background field as mentioned earlier. As in \cite{Tomsyukawa2} we now write \eqref{F1} in the form 
\begin{equation}
\Gamma_{\textrm{fermion}}=-i\,\ln\det\left( \begin{array}{cc} \mathbb{A}&0\\ 0&\mathbb{D}\end{array}\right) -i\,{\textrm{Tr}}\ln(\mathbb{I}+\mathbb{X}),\label{F3}
\end{equation}
where
\begin{equation}
\mathbb{X}=\left( \begin{array}{cc} 0&\mathbb{A}^{-1}\mathbb{B}\\ \mathbb{D}^{-1}\mathbb{C}&0\end{array}\right).\label{F4}
\end{equation}
Here we use ${\textrm{Tr}}$ to denote the functional as well as the trace over Dirac spinor and group indices. So for example, ${\textrm{Tr}}\,\mathbb{X}=\int dv_x\,{\textrm{tr}}\,\mathbb{X}(x,x)$ where ${\textrm{tr}}$ is just the normal Dirac and group trace. $\mathbb{A}^{-1}$ and $\mathbb{D}^{-1}$ are simply related to the Green functions $\Psi(x,x')$ and $\chi(x,x')$ for the $\Psi$ and $\chi$ spinors that we define by
\begin{align}
\Big(i\gamma^\mu \nabla_{\!\!\mu}-m_{\psi}-im_{\psi5}\gamma_5\Big)\Psi(x,x')=-\delta(x,x'),\label{F5}\\
\Big(i\gamma^\mu\nabla_{\!\!\mu}-m_{\chi}-im_{\chi5}\gamma_5\Big)\chi(x,x')=-\delta(x,x').\label{F6}
\end{align}
(So $\mathbb{A}^{-1}(x,x')=-\Psi(x,x')$ and $\mathbb{D}^{-1}(x,x')=-\chi(x,x')$.) 

The term in \eqref{F3} that involves ${\textrm{Tr}}\ln(\mathbb{I}+\mathbb{X})$ can be expanded in powers of $\mathbb{X}$. Because of the form of $\mathbb{X}$ in \eqref{F4} all terms that are odd in $\mathbb{X}$ will have a vanishing trace, and all terms that involve more than four factors of $\mathbb{X}$ will not have any poles in dimensional regularisation. This means that the pole part of $\Gamma_{\textrm{fermion}}$ is contained in the terms explicitly indicated below,
\begin{equation}
\Gamma_{\textrm{fermion}}=-i\,\ln\det\left( \begin{array}{cc} \mathbb{A}&0\\ 0&\mathbb{D}\end{array}\right) +\frac{i}{2}\,{\textrm{Tr}}(\mathbb{X}^2)+\frac{i}{4}\,{\textrm{Tr}}(\mathbb{X}^4)+\cdots.\label{F7}
\end{equation} 

Using \eqref{F4} it can be seen that
\begin{equation}
{\textrm{Tr}}(\mathbb{X}^2)=2\int dv_xdv_{x'}\,\textrm{tr}\lbrack\Psi(x,x')\,(w^\ast+iw_5^\ast\gamma_5)\,\chi(x',x)\,(w+iw_5\gamma_5)\rbrack\bar{\Phi}^\dagger(x)\bar{\Phi}(x'),\label{F8}
\end{equation}
and that
\begin{align}
{\textrm{Tr}}(\mathbb{X}^4)&=2\,\int dv_x\int dv_{x'}\int dv_{x''}\int dv_{x'''}\,\textrm{tr}\lbrack(w+iw_5\gamma_5)\Psi(x,x')(w^\ast+iw_5^\ast\gamma_5)\label{F9}\\
&\times\chi(x',x'') (w+iw_5\gamma_5)\Psi(x'',x''')(w^\ast+iw_5^\ast\gamma_5)\chi(x''',x)\rbrack\,\bar{\Phi}^\dagger(x'')\bar{\Phi}(x')\,\bar{\Phi}^\dagger(x) \bar{\Phi}(x''').\nonumber
\end{align}
The pole parts of the expressions needed in \eqref{F8} and \eqref{F9} are identical to those evaluated in \cite{Tomsyukawa2} so we will not repeat the details here.

The first term in \eqref{F7} is independent of the background scalar field and involves just the vacuum gravitational pole terms. It was shown in \cite{Tomsyukawa1,Tomsyukawa2} that
\begin{equation}
\ln\det(i\nablaslash-m_0-im_5\gamma_5)=\frac{1}{2}\,\ln\det(D^2+X),\label{F10}
\end{equation}
where
\begin{subequations}
\begin{align}
D_\mu&=\nabla_\mu-m_5\gamma_5\gamma_\mu,\label{F11a}\\
X&=\Big(m_0^2+3\,m_5^2+\frac{1}{4}\,R\Big)\,I+2i\,m_0\,m_5\,\gamma_5,\label{F11b}\\
W_{\mu\nu}&=\lbrack D_\mu,D_\nu\rbrack=-\,\frac{1}{4}\,R_{\mu\nu\lambda\sigma}\,\gamma^\lambda\gamma^\sigma-m_5^2\,\lbrack\gamma_\mu,\gamma_\nu\rbrack.\label{F11c}
\end{align}
\end{subequations}
This enables the standard heat kernel result \eqref{4.4.4} and \eqref{4.4.5} to be used for each of the two operators that arise from \eqref{F7} with \eqref{F2a} and \eqref{F2d}.

The net result for the pole part of $\Gamma_{\textrm{fermion}}$ is 
\begin{align}
{\textrm{PP}}\lbrace\Gamma_{\textrm{fermion}}\rbrace&=\frac{1}{4\pi^2\epsilon}\,\int dv_x\Big\lbrace  \frac{(d_R+1)}{288}\,R^2-\frac{(d_R+1)}{180}\,R_{\mu\nu}R^{\mu\nu}-\frac{7(d_R+1)}{1440}\,R_{\mu\nu\lambda\sigma}R^{\mu\nu\lambda\sigma}\nonumber\\
&+\frac{1}{12}\lbrack\,d_R\,(m_\psi^2+m_{\psi5}^2)+m_\chi^2+m_{\chi5}^2\rbrack\,R + \frac{d_R}{2}\,(m_\psi^2+m_{\psi5}^2)^2+\frac{1}{2}\,(m_\chi^2+m_{\chi5}^2)^2\nonumber\\
&-(|w|^2+|w_5|^2)\nabla^\mu\bar{\Phi}^\dagger\nabla_\mu\bar{\Phi}+\frac{1}{6}\,R|\bar{\Phi}|^2\nonumber\\
&+2\,\Big\lbrack (|w|^2+|w_5|^2)(m_\psi^2+m_{\psi5}^2+m_\chi^2+m_{\chi5}^2)\nonumber\\
&\qquad+(wm_\psi+w_5 m_{\chi5})(w^\ast m_\psi+w_5^\ast m_{\psi5})\nonumber\\
&\qquad-(wm_{\chi5}-w_5 m_{\chi})(w^\ast m_{\psi5}-w_5^\ast m_{\psi})\Big\rbrack\,|\bar{\Phi}|^2\nonumber\\
&+\Big\lbrack (|w|^2+|w_5|^2)^2-(w w_5^\ast-w^\ast w_5)^2\Big\rbrack\,|\bar{\Phi}|^4\Big\rbrace.\label{F12}
\end{align}

\section{One-loop counterterms and renormalisation group functions}
\label{sec-ct}

We will now combine the results for the complete pole parts of the one-loop effective action coming from Secs.~\ref{secBose} and \ref{secspinors} and then evaluate the one-loop counterterms and associated renormalisation group functions. Because only the scalar field background is kept non-zero in our calculation (apart from the arbitrary gravitational background) we can set the gauge fields to zero in \eqref{2.1}. The bare quantities in \eqref{2.1} and \eqref{2.5} can be expressed in terms of renormalised ones and counterterms in dimensional regularisation as originally specified by `t~Hooft~\cite{tHooft1973}:
\begin{subequations}\label{6.1}
\begin{align}
\Phi_\bare&=\mu^{\epsilon/2}(1+\delta Z_\varphi)\Phi,\label{6.1a}\\
m_{s\,\bare}^2&=m_s^2+\delta m_s^2,\label{6.1b}\\
\xi_\bare&=\xi+\delta\xi,\label{6.1c}\\
\lambda_\bare&=\mu^{-\epsilon}(\lambda+\delta\lambda),\label{6.1d}\\
\Lambda_{\bare}&=\mu^{\epsilon}(\Lambda+\delta\Lambda),\label{6.1e}\\
\kappa_{\bare}&=\mu^{\epsilon}(\kappa+\delta\kappa),\label{6.1f}\\
\alpha_{i\,\bare}&=\mu^{\epsilon}(\alpha_i+\delta\alpha_i) \ {\textrm{for}}\ i=1,2,3.\label{6.1g}
\end{align}
\end{subequations}
Here $\mu$ is the `t~Hooft arbitrary unit of mass introduced so that all of the renormalised quantities (those without the subscript `B' appended) have the same dimensions, in units of mass, for all spacetime dimensions $n=4+\epsilon$ as they do for $n=4$. Additional counterterms and poles will arise in the effective action for the full theory with all fields having a non-zero background part but these extra terms are all irrelevant for the scalar field part that we are considering here. 

The counterterm part of the classical action, which will cancel the pole terms coming from the one-loop part of the effective action, follow from \eqref{2.1} and \eqref{2.5} using \eqref{6.1} and is
\begin{align}
S_{\textrm{ct}}&=\int dv_x\Big\lbrack 2\,\delta Z_\varphi \nabla^\mu\bar{\Phi} \nabla_\mu\bar{\Phi} - (\delta m_s^2+2\, \delta Z_\varphi\,m_s^2)|\bar{\Phi}|^2- (\delta \xi+2\, \delta Z_\varphi\,\xi)\,R\,|\bar{\Phi}|^2\label{6.2}\\
&\quad-\left(\frac{\delta\lambda}{6}+\frac{2}{3}\,\lambda\,\delta Z_\varphi\right)\,|\bar{\Phi}|^4+\delta\Lambda+\delta\kappa R+\delta\alpha_1 R_{\mu\nu\rho\sigma}R^{\mu\nu\rho\sigma} +\delta\alpha_2 R_{\mu\nu}R^{\mu\nu}+\delta\alpha_3 R^2\Big\rbrack.\nonumber
\end{align}
The pole terms that come in the one-loop effective action by combining \eqref{4.6.1} and \eqref{F12} are all of the form of the counterterms in \eqref{6.2}, so all poles can be dealt with with the indicated counterterms. Requiring that the poles from the one-loop effective action be cancelled by the terms in \eqref{6.2} leads to
\begin{subequations}\label{6.3}
\begin{align}
\delta Z_\varphi&=\frac{1}{16\pi^2\epsilon}\left( 2\,|w|^2+2\,|w_5|^2-3\,e_s^2-3\,C_2\,g^2-2\,\zeta_1e_s-2\,\zeta_2\,C_2\,g\right),\label{6.3a}\\
\delta m_s^2&=\frac{1}{8\pi^2\epsilon}\Big\lbrace \Big\lbrack 3\,e_s^2+3\,C_2\,g^2-(d_R+1)\,\frac{\lambda}{3}-2\,|w|^2-2\,|w_5|^2\Big\rbrack\,m_s^2\nonumber\\
&\qquad+4(|w|^2+|w_5|^2)(m_\psi^2+m_{\psi5}^2+m_\chi^2+m_{\chi5}^2)\nonumber\\
&\qquad+4(wm_\psi+w_5 m_{\chi5})(w^\ast m_\psi+w_5^\ast m_{\psi5})\nonumber\\
&\qquad-4(wm_{\chi5}-w_5 m_{\chi})(w^\ast m_{\psi5}-w_5^\ast m_{\psi})\Big\rbrace,\label{6.3b}\\
\delta \xi&=\frac{1}{8\pi^2\epsilon}\Big\lbrack 3\,e_s^2+3\,C_2\,g^2-(d_R+1)\,\frac{\lambda}{3}-2\,|w|^2-2\,|w_5|^2\Big\rbrack\,\left(\xi-\frac{1}{6}\right),\label{6.3c}\\
\delta\lambda&=-\frac{1}{4\pi^2\epsilon}\Big\lbrace 9\,e_s^4-3\,e_s^2\lambda-3\,C_2\,g^2\,\lambda+18\,C_2\Big(\frac{d_R-1}{d_G}\Big)\,e_s^2\,g^2+\frac{1}{6}(d_R+1)\,\lambda^2\nonumber\\
&\quad+9\,C_2^2\,\Big\lbrack\frac{(d_R-1)(d_R^2+2\,d_R-2)}{2\,d_G^2}\Big\rbrack\,g^4+6\,(ww_5^\ast-w^\ast w_5)^2-6\,(|w|^2+|w_5|^2)^2\Big\rbrace,\label{6.3d}\\
\delta\Lambda&=-\frac{1}{16\pi^2\epsilon} \Big\lbrack 2\,(m_\chi^2+m_{\chi5}^2)^2+2\,d_R\,(m_\psi^2+m_{\psi5}^2)^2 -d_R\,m_s^4\Big\rbrack,\label{6.3e}\\
\delta\kappa&=-\frac{1}{48\pi^2\epsilon} \Big\lbrack m_\chi^2+m_{\chi5}^2+d_R\,(m_\psi^2+m_{\psi5}^2) -6\,d_R\,\Big(\xi-\frac{1}{6}\Big)\,m_s^2\Big\rbrack,\label{6.3f}\\
\delta\alpha_1&=-\frac{1}{5760\pi^2\epsilon}(15\,d_R+19),\label{6.3g}\\
\delta\alpha_2&=\frac{1}{1440\pi^2\epsilon}(45\,d_R+46),\label{6.3h}\\
\delta\alpha_3&=-\frac{1}{1152\pi^2\epsilon}\Big\lbrack11\,d_R+11-72\,d_R\,\Big(\xi-\frac{1}{6}\Big)^2\Big\rbrack.\label{6.3i}
\end{align}
\end{subequations}

In the standard background field method the two gauge parameters $\zeta_1$ and $\zeta_2$ that we introduced in the gauge conditions \eqref{3.4} are completely arbitrary. From the counterterms calculations above it can be seen that the only place that they enter is in the scalar field renormalisation \eqref{6.3a}. All other counterterms are completely independent of these arbitrary parameters as they cancel out from the pole terms in the one-loop effective action from whence they came. This would not be the case if the scalar field renormalisation was erroneously ignored by simply taking a constant scalar field background from the start. As we emphasised in Sec.~\ref{secbfmethod}, the correct expression for the effective action is obtained by taking $\zeta_1=e_s$ and $\zeta_2=g$ to implement the Landau-DeWitt gauge as noted in \eqref{3.5}. The correct gauge condition independent field renormalisation factor is therefore given by
\begin{equation}\label{6.4}
\delta Z_\varphi=-\frac{1}{16\pi^2\epsilon}\left( 5\,e_s^2+5\,C_2\,g^2 -2\,|w|^2-2\,|w_5|^2\right).
\end{equation}
What this calculation has shown, by keeping the gauge parameters arbitrary, is that all of the counterterms for the scalar sector of the gauge theory considered here are independent of the gauge parameters apart from the field renormalisation factor; the Vilkovisky-DeWitt corrections to the usual background field method are only needed to calculate the scalar field renormalisation, at least for gauge conditions of the form \eqref{3.4}.

The one-loop renormalisation group functions now follow directly from the one-loop counterterms \eqref{6.3} using the pioneering method of `t~Hooft~\cite{tHooft1973}. We will follow the review in \cite{ParkerTomsbook} where if $q^i$ represents any of the renormalised expressions $(\bar{\Phi},m_s^2,\xi,\lambda,\Lambda,\kappa,\alpha_1,\alpha_2,\alpha_3)$, then under a change of the renormalisation mass $\mu$
\begin{equation}\label{6.5}
\mu\,\frac{d}{d\mu}\,q^i=\beta_{q^i}.
\end{equation}
$\beta_{q^i}$ are the renormalisation group functions. The explicit results that follow from \eqref{6.3b}--\eqref{6.3i} and \eqref{6.4} are
\begin{subequations}\label{6.6}
\begin{align}
\beta_{\bar{\Phi}}&=\frac{1}{16\pi^2}\left( 5\,e_s^2+5\,C_2\,g^2 -2\,|w|^2-2\,|w_5|^2\right)\,\bar{\Phi},\label{6.6a}\\
\beta_{m_s^2}&=-\frac{1}{8\pi^2}\Big\lbrace \Big\lbrack 3\,e_s^2+3\,C_2\,g^2-(d_R+1)\,\frac{\lambda}{3}-2\,|w|^2-2\,|w_5|^2\Big\rbrack\,m_s^2\nonumber\\
&\qquad+4(|w|^2+|w_5|^2)(m_\psi^2+m_{\psi5}^2+m_\chi^2+m_{\chi5}^2)\nonumber\\
&\qquad+4(wm_\psi+w_5 m_{\chi5})(w^\ast m_\psi+w_5^\ast m_{\psi5})\nonumber\\
&\qquad-4(wm_{\chi5}-w_5 m_{\chi})(w^\ast m_{\psi5}-w_5^\ast m_{\psi})\Big\rbrace,\label{6.6b}\\
\beta_{\xi}&=-\frac{1}{8\pi^2}\Big\lbrack 3\,e_s^2+3\,C_2\,g^2-(d_R+1)\,\frac{\lambda}{3}-2\,|w|^2-2\,|w_5|^2\Big\rbrack\,\left(\xi-\frac{1}{6}\right),\label{6.6c}\\
\beta_{\lambda}&=\frac{1}{4\pi^2}\Big\lbrace 9\,e_s^4-3\,e_s^2\lambda-3\,C_2\,g^2\,\lambda+18\,C_2\Big(\frac{d_R-1}{d_G}\Big)\,e_s^2\,g^2+\frac{1}{6}(d_R+1)\,\lambda^2\label{6.6d}\\
&\quad+9\,C_2^2\,\Big\lbrack\frac{(d_R-1)(d_R^2+2\,d_R-2)}{2\,d_G^2}\Big\rbrack\,g^4+6\,(ww_5^\ast-w^\ast w_5)^2-6\,(|w|^2+|w_5|^2)^2\Big\rbrace,\nonumber\\
\beta_{\Lambda}&=\frac{1}{16\pi^2} \Big\lbrack 2\,(m_\chi^2+m_{\chi5}^2)^2+2\,d_R\,(m_\psi^2+m_{\psi5}^2)^2 -d_R\,m_s^4\Big\rbrack,\label{6.6e}\\
\beta_{\kappa}&=\frac{1}{48\pi^2} \Big\lbrack m_\chi^2+m_{\chi5}^2+d_R\,(m_\psi^2+m_{\psi5}^2) -6\,d_R\,\Big(\xi-\frac{1}{6}\Big)\,m_s^2\Big\rbrack,\label{6.6f}\\
\beta_{\alpha_1}&=\frac{1}{5760\pi^2}(15\,d_R+19),\label{6.6g}\\
\beta_{\alpha_2}&=-\frac{1}{1440\pi^2}(45\,d_R+46), \label{6.6h}\\
\beta_{\alpha_3}&=\frac{1}{1152\pi^2}\Big\lbrack11\,d_R+11-72\,d_R\,\Big(\xi-\frac{1}{6}\Big)^2\Big\rbrack.\label{6.6i}
\end{align}
\end{subequations}
We will use these results in the next section to consider the leading terms in the renormalised one-loop effective action.

\section{Effective action}\label{effectivepotential}

The renormalisation group can be used to obtain terms in the effective action as discussed originally by Coleman and Weinberg~\cite{ColemanandWeinberg} in flat spacetime and by Buchbinder and Odintsov~\cite{BuchbinderandOdintsovRG} in curved spacetime. We will only look at the simplest case here where there are no mass scales present in the theory (apart from the fields), so we will take $m_s^2=m_\chi=m_{\chi5}=m_\psi=m_{\psi5}=0$ in our previous expressions resulting in great simplifications for the renormalisation group functions in \eqref{6.6}.

As described in \cite{Tomsyukawa1} the general form of the effective action, including only terms that follow from the renormalisation group, is
\begin{align}
\Gamma&=\int dv_x\Big\lbrack \frac{1}{2}Z(\bar{\Phi})\partial^\mu\bar{\Phi}\partial_\mu\bar{\Phi}-V_0(\bar{\Phi})-RV_1(\bar{\Phi})\nonumber\\
&\qquad+\alpha_{1}(\bar{\Phi})R^{\mu\nu\lambda\sigma}R_{\mu\nu\lambda\sigma} + \alpha_{2}(\bar{\Phi})R^{\mu\nu}R_{\mu\nu} + \alpha_{3}(\bar{\Phi})R^2 \Big\rbrack,\label{7.1}
\end{align}
where $Z(\bar{\Phi}),V_0(\bar{\Phi}),V_1(\bar{\Phi}),\alpha_i(\bar{\Phi})$ are to be determined by requiring that the effective action be invariant under a rescaling of the renormalisation mass $\mu$. The Coleman-Weinberg renormalisation conditions on the potentials appearing in \eqref{7.1} are given by
\begin{subequations}\label{7.2}
\begin{align}
Z(\bar{\Phi}=\mu)&=1,\label{7.2a}\\
\left.\frac{\partial^2 V_0(\bar{\Phi})}{\partial^2|\bar{\Phi}|^2}\right|_{\bar{\Phi}=\mu}&=\frac{\lambda}{3},\label{7.2b}\\
\left.\frac{\partial V_1(\bar{\Phi})}{\partial|\bar{\Phi}|^2}\right|_{\bar{\Phi}=\mu}&=\xi,\label{7.2c}\\
\alpha_i(\bar{\Phi}=\mu)&=\alpha_i.\label{7.2d}
\end{align}
\end{subequations}
With these renormalisation conditions the general form of the potentials can be written in the form
\begin{subequations}\label{7.3}
\begin{align}
Z(\bar{\Phi})&=1+A\,\ln\left(|\bar{\Phi}|^2/\mu^2\right),\label{7.3a}\\
V_0(\bar{\Phi})&=\frac{\lambda}{6}\,|\bar{\Phi}|^4+B\,|\bar{\Phi}|^4\left\lbrack\ln\left(|\bar{\Phi}|^2/\mu^2\right)-\frac{3}{2}\right\rbrack,\label{7.3b}\\
V_1(\bar{\Phi})&=\xi\,|\bar{\Phi}|^2+C\,|\bar{\Phi}|^2\left\lbrack\ln\left(|\bar{\Phi}|^2/\mu^2\right)-1\right\rbrack,\label{7.3c}\\
\alpha_i(\bar{\Phi})&=\alpha_i+D_i\,\ln\left(|\bar{\Phi}|^2/\mu^2\right).\label{7.3d}
\end{align}
\end{subequations}
Here $A,B,C,D_i$ are constants, independent of the background field, that are determined by the one-loop renormalisation group functions given in \eqref{6.6}. Requiring that the effective action in \eqref{7.1} with the potentials taking the form of \eqref{7.3} be invariant under a rescaling of the renormalisation mass $\mu$ results in
\begin{subequations}\label{7.4}
\begin{align}
A&=\tilde{\beta}_{\bar{\Phi}},\label{7.4a}\\
B&=\frac{1}{12}\,\beta_\lambda+\frac{1}{3}\,\lambda\,\tilde{\beta}_{\bar{\Phi}},\label{7.4b}\\
C&=\frac{1}{2}\,\beta_\xi+\xi\,\tilde{\beta}_{\bar{\Phi}},\label{7.4c}\\
D_i&=\beta_{\alpha_i}.\label{7.4d}
\end{align}
\end{subequations}
where
\begin{equation}\label{7.5}
\tilde{\beta}_{\bar{\Phi}}=\frac{1}{16\pi^2}\left( 5\,e_s^2+5\,C_2\,g^2 -2\,|w|^2-2\,|w_5|^2\right),
\end{equation}
is the coefficient multiplying $\bar{\Phi}$ in \eqref{6.6a}. As already remarked above, this expression would contain the gauge parameters $\zeta_1$ and $\zeta_2$ if calculated using conventional methods (as found from \eqref{6.3a}) and lead to the erroneous conclusion that the potentials were gauge condition dependent. We will not write out the explicit forms for the coefficients in \eqref{7.4a}--\eqref{7.5} as they all follow readily from the renormalisation group functions.

It is also worth commenting on the anomaly that was remarked on in \cite{Tomsyukawa1,Tomsyukawa2} when removing the pseudoscalar mass term, or else the pseudoscalar Yukawa coupling by a field redefinition. We can define two new fields, $\chi^\prime$ and $\Psi^\prime$ by
\begin{align}
\chi(x)&=e^{-i\theta\gamma_5}\,\chi^\prime(x),\label{7.6a}\\
\Psi(x)&=e^{-i\omega\gamma_5}\,\Psi^\prime(x),\label{7.6b}
\end{align}
where the angles $\theta$ and $\omega$ are chosen to eliminate the pseudoscalar mass terms (or the pseudoscalar Yukawa terms if desired) in \eqref{2.4}. The pseudoscalar mass terms can be removed by choosing
\begin{align}
\theta&=\frac{1}{2}\,\tan^{-1}\left(\frac{m_{\chi5}}{m_\chi}\right),\label{7.7a}\\
\omega&=\frac{1}{2}\,\tan^{-1}\left(\frac{m_{\psi5}}{m_\psi}\right).\label{7.7b}
\end{align}
(Similar results apply with the Yukawa couplings appearing in place of the masses if the pseudoscalar Yukawa couplings are eliminated.) The transformations \eqref{7.6a} and \eqref{7.6b} will also transform the Yukawa coupling terms in an easily determined way. It can be shown that the one-loop counterterms that were obtained above, and the one loop renormalisation group functions that follow from them, are invariant under the change of coupling constants and masses induced by the transformations in \eqref{7.6a} and \eqref{7.6b} as in \cite{Tomsyukawa1,Tomsyukawa2}. The classical theory based on $\chi,\Psi$ and the one based on $\chi^\prime,\Psi^\prime$ are completely equivalent, but this is not necessarily the case for the quantum theories. The technical details of examining the potential anomaly are virtually identical to those involved with the chiral or axial anomaly which is hardly surprising given the form of the transformations. The details are most easily calculated by examiing the change in the functional measure under \eqref{7.6a} and \eqref{7.6b} following the treatment initiated by Fujikawa~\cite{Fujikawa1,Fujikawa2,Fujikawa3}. The details in the calculation needed here are outlined in \cite[Appendix~C]{Tomsyukawa2} and need not be repeated. If we call $\Gamma$ the effective action for the theory that uses the original fields $\chi,\Psi$ and $\Gamma^\prime$ the effective action for the theory based on the fields $\chi^\prime,\Psi^\prime$ then the Fujikawa analysis results in
\begin{equation}
\Gamma^\prime=\Gamma-\frac{i\theta}{8\pi^2}\int dv_x\,\tr\lbrack\gamma_5\,E_2\rbrack-\frac{i\omega}{8\pi^2}\int dv_x\,\tr\lbrack\gamma_5\,\tilde{E}_2\rbrack.\label{7.8}
\end{equation}
Here $E_2$ is the heat kernel coefficient for the Dirac operator for the $\chi$ field that follows from \eqref{2.4} and \eqref{2.4a}, and $\tilde{E}_2$ is the heat kernel coefficient for the Dirac operator for the $\Psi$ field that follows from \eqref{2.4} and \eqref{2.4b}. The $E_2$ expressions can be found from \eqref{4.4.4} and \eqref{4.4.5}. The trace in \eqref{7.8} is over both Dirac indices and group indices if present. After a bit of calculation it can be shown that \eqref{7.8} becomes
\begin{align}
\Gamma^\prime&=\Gamma-\frac{(\theta+d_R\,\omega)}{384\pi^2}\int dv_x\,\epsilon^{\mu\nu\lambda\sigma}R_{\alpha\beta\mu\nu}R^{\alpha\beta}{}_{\lambda\sigma} + \frac{(e_\chi^2\,\theta+d_R\,e_\psi^2\,\omega)}{16\pi^2}\int dv_x\,\epsilon^{\mu\nu\lambda\sigma}F_{\mu\nu}F_{\lambda\sigma}\nonumber\\
&\qquad+\frac{g^2\,d_R\,\omega}{16\pi^2\,d_G}\,C_2\,\int dv_x\,\epsilon^{\mu\nu\lambda\sigma}B_{A\,\mu\nu}B_{A\,\lambda\sigma},\label{7.9}
\end{align}
if we leave in potential background gauge fields. The explicit expressions for the angles $\theta$ and $\omega$ are given in \eqref{7.7a} and \eqref{7.7b}. The result in \eqref{7.9} is consistent with the invariance of the one-loop counterterms under the required field redefinitions and shows that the two quantum theories are not identical, but are related by an anomaly that does not affect the divergent part of the effective action.

\section{Conclusions}\label{secconclusions}

The renormalisation group functions for a general gauge theory based on the gauge group $G\times U(1)$ to one loop order in a theory that contains scalars, Dirac spinors and gauge fields were calculated for the scalar sector of the theory. These renormalisation group functions were then used to obtain terms in the one-loop effective action. The results were obtained in a gauge condition independent way and we showed explicitly where the gauge conditions entered the more standard background field method. This dependence was traced through the renormalisation group functions and the correct gauge independent results were found. The field transformations which can be used to remove the pseudoscalar mass term for the Dirac fields in the classical theory was shown to lead to an anomaly in the quantum theory. This anomaly does not affect the divergent part of the effective action and therefore does not affect the renormalisation group functions that have been calculated. 

The calculation falls short of a full one-loop analysis as only the background scalar field was kept non-zero. There is no impediment to obtaining results for the complete quantised theory, other than technical difficulty.

\appendix\section{Some conventions}\label{appendA}

We adopt mostly the same conventions as in \cite{ParkerTomsbook}. Specifically, the spacetime metric has signature $-2$, the Riemann tensor is defined in terms of the Christoffel symbols by
\begin{equation}
R^{\lambda}{}_{\tau\mu\nu}=\Gamma^{\lambda}_{\tau\mu,\nu}- \Gamma^{\lambda}_{\tau\nu,\mu} + \Gamma^{\lambda}_{\nu\sigma}\Gamma^{\sigma}_{\mu\tau} - \Gamma^{\lambda}_{\mu\sigma}\Gamma^{\sigma}_{\nu\tau},\label{A1}
\end{equation}
and the Ricci tensor is $R_{\mu\nu}=R^{\lambda}{}_{\mu\lambda\nu}$.

The non-Abelian part of the gauge group, $G$, is assumed to be simple with its Lie algebra defined in terms of Hermitian generators $T_A$ by
\begin{equation}
\lbrack T_A,T_B\rbrack=i\,f_{ABC}\,T_C.\label{A3}
\end{equation}
Repeated group indices are labelled by $A,B,C,\ldots$ and run over $1,\ldots,d_G$ where $d_G$ is the dimension of the group. The structure constants $f_{ABC}$ are assumed to be totally antisymmetric.

If we have some particular representation of $G$, say $G_R$, then $T_A$ can be viewed as $d_R\times d_R$ Hermitian matrices, where $d_R$ gives the dimension of the representation $G_R$. The quadratic Casimir invariant is defined by
\begin{equation}
T_AT_A=C_2(G_R)\,I,\label{A4}
\end{equation}
for any representation $G_R$ with $I$ the $d_R\times d_R$ identity matrix. Because $G$ is assumed to be simple (we have explicitly extracted a $U(1)$ factor in the overall gauge group) and compact, we have
\begin{equation}
\tr(T_A)=0.\label{A5}
\end{equation}
We also have
\begin{equation}
\tr(T_AT_B)=\frac{d_R}{d_G}\,C_2(G_R)\,\delta_{AB},\label{A6}
\end{equation}
where the constant of proportionality is fixed by consistency with \eqref{A4}. It is customary in particle physics applications to fix the normalisation of $T_A$ and the structure constants in \eqref{A3} such that the right hand side of \eqref{A6} becomes $\frac{1}{2}\delta_{AB}$ but we do not do this here. The right hand side of \eqref{A6} does become $\frac{1}{2}\delta_{AB}$ for the special case of the fundamental representation of $SU(n)$.

Another important result is that for any representation $G_R$ of the simple group $G$ we have
\begin{equation}
(T_A)_{ij}(T_A)_{kl}=\frac{d_R}{d_G}\,C_2(G_R)\,\left(\delta_{il}\delta_{jk}-\frac{1}{d_R}\,\delta_{ij}\delta_{kl} \right),\label{A9}
\end{equation}
where $(T_A)_{ij}$ denotes the matrix elements of the generators $T_A$ in the representation $G_R$. This result is easily proven from the fact that any $d_R\times d_R$ Hermitian matrix can be expressed as a linear combination of the Lie algebra generators $T_A$ along with the identity matrix. The result in \eqref{A9} allows terms like $(\bphid T_A\bphi)^2$ and $(\bphid T_AT_B\bphi)^2$ to be related to $|\bphi|^4$ along with some group theoretic factors. Specifically, we will define
\begin{equation}
\tau_A^2=(\bphid T_A\bphi)^2=\frac{(d_R-1)}{d_G}\,C_2(G_R)\,|\bphi|^4,\label{A10}
\end{equation}
and
\begin{equation}
\rho_{AB}^2=(\bphid T_{(A}T_{B)}\bphi)^2=\frac{(d_R-1)(d_R^2+2\,d_R-2)}{2\,d_G^2}\,C_2^2(G_R)\,|\bphi|^4,\label{A11}
\end{equation}
where we have adopted the convention $2\,T_{(A}T_{B)}=T_AT_B+T_BT_A$. The expressions in \eqref{A10} and \eqref{A11} allow all of the pole terms in the effective action that are quartic in $\bphi$ to be related to the manifestly gauge invariant expression $|\bphi|^4$ and to be removed with the $\delta\lambda$ counterterm.

\section{Some Green's function results}\label{appGreen}

Let $D_{\mu\nu}(x,x')$ satisfy the equation
\begin{equation}
\Big\lbrack g_{\mu\lambda}\Box+R_{\mu\nu}-\Big(1-\frac{1}{\alpha}\Big)\nabla_\mu\nabla_\lambda\Big\rbrack\,D^{\lambda\nu}(x,x')=\delta^{\nu}_{\mu}\delta(x,x').\label{4.32}
\end{equation}
This is the vector field Green's function for the non-minimal operator that was used in \cite{Tomsyukawa2}. We will show that $\widetilde{G}^{\mu\nu}(x,x')$ defined in \eqref{4.26} coincides with $D^{\mu\nu}(x,x')$ in the Landau gauge, $\alpha\rightarrow0$. This is simple to show in flat spacetime using Fourier transforms, but slightly more involved in curved spacetime. The fact that we need to use $\widetilde{G}^{\mu\nu}(x,x')$ is where non-minimal operators, although initially absent from the calculations, come back in. The appearance of non-minimal operators appears to be unavoidable.

With $G_{\mu\nu}(x,x')$ defined in \eqref{4.18} it follows from \eqref{4.32} that
\begin{equation}
D_{\mu\nu}(x,x')=G_{\mu\nu}(x,x')-\left(1-\frac{1}{\alpha}\right)\int dv_{x''}\nabla^{\prime\prime\sigma}D_{\nu\sigma}(x,x'')\nabla^{\prime\prime\rho}G_{\rho\nu}(x'',x').\label{4.33}
\end{equation}
Note that by comparing \eqref{4.32} with \eqref{4.18} it is clear that
\begin{equation}
G_{\mu\nu}(x,x')=\left. D_{\mu\nu}(x,x')\right|_{\alpha=1}.\label{4.34}
\end{equation}
\eqref{4.33} is consistent with this relation.

By operating on both sides of \eqref{4.32} it can be seen that
\begin{equation}
\nabla^\lambda D_{\lambda\nu}(x,x')=-\alpha\,\nabla^\prime_\nu D(x,x'),\label{4.35}
\end{equation}
where
\begin{equation}
\Box\,D(x,x')=\delta(x,x').\label{4.36}
\end{equation}
Similarly,
\begin{equation}
\nabla^{\prime\nu}D_{\mu\nu}(x,x')=-\alpha\,\nabla_\mu D(x,x').\label{4.37}
\end{equation}
By using \eqref{4.34}, \eqref{4.35} and \eqref{4.37} in \eqref{4.33} if follows that
\begin{equation}
D_{\mu\nu}(x,x')=G_{\mu\nu}(x,x')+(1-\alpha)\nabla_\mu\nabla^\prime_\nu\int dv_{x''}D(x,x'')D(x'',x').\label{4.38}
\end{equation}
From \eqref{4.26} if we set $\alpha=1$ in \eqref{4.35} and \eqref{4.37} we see that
\begin{equation}
\widetilde{G}_{\mu\nu}(x,x')=G_{\mu\nu}(x,x')+\nabla_\mu\nabla^\prime_\nu\int dv_{x''}D(x,x'')D(x'',x').\label{4.39}
\end{equation}
Comparison of \eqref{4.38} and \eqref{4.39} shows that
\begin{equation}
\widetilde{G}_{\mu\nu}(x,x')=\left. D_{\mu\nu}(x,x')\right|_{\alpha=0},\label{4.40}
\end{equation}
as we claimed. This allows us to use our previous results~\cite{Tomsyukawa2} for the local momentum space expansions of the vector field Green's functions and simply set $\alpha=0$ there.

If we refer the spacetime labels back to a local orthonormal frame using the vierbein formalism\footnote{The reason why this is done, as explained in \cite{toms2014local} is that it means we do not need to consider the bivector of geodetic parallel displacement as in \cite{DeWittdynamical}.} then the results of \cite{Tomsyukawa2} give (where we use $a,b,\ldots$ to represent orthonormal frame indices)
\begin{align}
G^{a}{}_{b}(x,x')&=\int\frac{d^np}{(2\pi)^n}\,e^{ip\cdot y}\Big\lbrack -\delta^{a}_{b}\,p^{-2} + \frac{1}{3}\,\delta^{a}_{b}\,R\,p^{-4}-R^{a}{}_{b}p^{-4}\nonumber\\
&\quad -\frac{2}{3}\,\delta^{a}_{b}\,R^{\mu\nu}\,p_\mu p_\nu\,p^{-6}+\cdots\Big\rbrack,\label{4.41}
\end{align}
(by taking $\alpha=1$ in the relevant expressions of \cite{Tomsyukawa2}) and
\begin{align}
\widetilde{G}^{a}{}_{b}(x,x')&=\int\frac{d^np}{(2\pi)^n}\,e^{ip\cdot y}\Big\lbrack -\delta^{a}_{b}\,p^{-2} + p^ap_bp^{-4}+ \frac{1}{3}\,\delta^{a}_{b}\,R\,p^{-4} -\frac{2}{3}\,R\,p^ap_b\,p^{-6}
- \frac{7}{6}\,R^{a}{}_{b}\,p^{-4}\nonumber\\
&\quad -\frac{2}{3}\,\delta^{a}_{b}\,R^{\mu\nu}\,p_\mu p_\nu\,p^{-6}+2\,R^{\mu\nu}\,p_\mu p_\nu p^a p_b\,p^{-8}
+\frac{2}{3}\,R^{a\mu}{}_{b}{}^{\nu}\,p_\mu p_\nu\,p^{-6}+\cdots\Big\rbrack,\label{4.42}
\end{align}
(by taking $\alpha=0$ in the relevant expressions of \cite{Tomsyukawa2}). We have only indicated the terms in the local momentum space expansion that are linear in the curvature; higher order terms can be found as described in \cite{toms2014local}. The scalar field Green's function expansion is
\begin{align}
\Delta(x,x')&= \int\frac{d^np}{(2\pi)^n}\,e^{ip\cdot y}\Big\lbrack  \frac{1}{p^2} + \left(\xi-\frac{1}{3}\right)\,R\,p^{-4}+m_s^2\,p^{-4}+\frac{2}{3}\,R^{\mu\nu}\,p_\mu p_\nu\, p^{-6}+\cdots\Big\rbrack.\label{4.43}
\end{align}

We summarise for convenience some of the basic results needed in our calculation. Full details can be found in \cite{Tomsyukawa2}. The basic results of dimensional regularisation~\cite{tHooftandVeltman} are
\begin{subequations}\label{B1}
\begin{align}
{\textrm{PP}}\left\lbrace \int\frac{d^np}{(2\pi)^n}\,\frac{1}{p^4} \right\rbrace&=-\frac{i}{8\pi^2\epsilon},\label{B1a}\\
{\textrm{PP}}\left\lbrace \int\frac{d^np}{(2\pi)^n}\,\frac{p_\mu p_\nu}{p^6} \right\rbrace&=-\frac{i}{32\pi^2\epsilon}\,\eta_{\mu\nu},\label{B1b}\\
{\textrm{PP}}\left\lbrace \int\frac{d^np}{(2\pi)^n}\,\frac{p_\mu p_\nu p_\lambda p_\sigma}{p^8} \right\rbrace&=-\frac{i}{192\pi^2\epsilon}\,(\eta_{\mu\nu}\eta_{\lambda\sigma} + \eta_{\mu\lambda}\eta_{\nu\sigma} + \eta_{\mu\sigma}\eta_{\lambda\nu}).\label{B1c}
\end{align}
\end{subequations}
With these used we find from \eqref{4.43} that
\begin{equation}
{\textrm{PP}}\left\lbrace \Delta(x,x) \right\rbrace=-\frac{i}{8\pi^2\epsilon}\Big\lbrack m_s^2+\Big(\xi-\frac{1}{6}\Big)R\Big\rbrack.\label{B2}
\end{equation}
and from \eqref{4.42} that
\begin{equation}
{\textrm{PP}}\left\lbrace \widetilde{G}^{\mu}{}_{\mu}(x,x) \right\rbrace=\frac{i}{16\pi^2\epsilon}\,R.\label{B3}
\end{equation}
Here ${\textrm{PP}}\lbrace\cdots\rbrace$ denotes the pole part of any expression. By taking the $\alpha\rightarrow0$ limits of results in \cite{Tomsyukawa2} we have
\begin{align}
{\textrm{PP}}\left\lbrace \Delta(x,x')\widetilde{G}^{\mu\nu}(x,x') \right\rbrace&=\frac{3i}{32\pi^2\epsilon}\,g^{\mu\nu}\delta(x,x'),\label{B4}\\
\textrm{PP}\left\lbrace G^{\mu}(x,x')\Delta(x,x') \right\rbrace&=-\frac{i}{16\pi^2\epsilon}\,\nabla^\mu\delta(x,x'),\label{B5}\\
\textrm{PP}\left\lbrace\Delta^2(x,x')\right\rbrace&=-\frac{i}{8\pi^2\epsilon}\,\delta(x,x'),\label{B6}\\
\textrm{PP}\left\lbrace \widetilde{G}^{\mu\nu}(x,x')\widetilde{G}_{\mu\nu}(x,x')\right\rbrace
&=-\frac{3i}{8\pi^2\epsilon}\,\delta(x,x').\label{B7}
\end{align}
In \eqref{B5} we have $G^\mu(x,x')=\nabla^\prime_\nu G^{\mu\nu}(x,x')$ as defined in \eqref{4.29}. Because the results of \eqref{B4}--\eqref{B7} do not involve the spacetime curvature they are easily checked with a flat spacetime calculation.

\section{Some details in the evaluation of $\Gamma_4$}\label{appGamma4}

Write $S_2$ in \eqref{4.8} in the generic form
\begin{equation}
S_2=S_{2\Phi}+S_{2G},\label{C1}
\end{equation}
where 
\begin{equation}
S_{2\Phi}=\int dv_x\Big\lbrack \alpha_{ij}(x)\,\Phi_i\Phi_j+\beta_{ij}(x)\,\Phi_i^\dagger\Phi_j^\dagger+\gamma_{ij}(x)\,\Phi_i^\dagger\Phi_j\Big\rbrack,\label{C2}
\end{equation}
involves just the quantum scalar fields, and
\begin{equation}
S_{2G}=\int dv_x\Big\lbrack \epsilon(x)\,A^\mu A_\mu + 2\,e_s\,g\,\tau_A(x)\,A^\mu B_{A\mu} + g^2\,\rho_{AB}(x)\,B_{A}^{\mu}B_{A\mu}\Big\rbrack,\label{C3}
\end{equation}
involves just the quantum gauge fields. Here $\alpha_{ij}, \beta_{ij}$ and $\gamma_{ij}$ were defined in \eqref{4.9}, $\tau_A$ was defined in \eqref{3.8a}, $\rho_{AB}$ was defined in \eqref{3.8}, and we have defined
\begin{equation}
\epsilon(x)=e_s^2\,|\bar{\Phi}(x)|^2.\label{C4}
\end{equation}

In a similar way we write the last two terms of $S_1$ in \eqref{4.7} as
\begin{equation}
S_1=S_{11}+S_{12}+S_{13}+S_{14},\label{C6}
\end{equation}
where 
\begin{subequations}\label{C7}
\begin{align}
S_{11}&= i\,\int dv_x\,\zeta_{1i}(x)\,\sigma(x)\Phi_i^\dagger(x),\label{C7a}\\
S_{12}&= -i\,\int dv_x\,\bar{\zeta}_{1i}(x)\,\sigma(x)\Phi_i(x),\label{C7b}\\
S_{13}&= i\,\int dv_x\,\zeta_{2Ai}(x)\,\sigma_A(x)\Phi_i^\dagger(x),\label{C7c}\\
S_{14}&= -i\,\int dv_x\,\bar{\zeta}_{2Ai}(x)\,\sigma_A(x)\Phi_i(x).\label{C7d}
\end{align}
\end{subequations}
Although we used a similar notation for different expressions in Sec.~\ref{secquad} this should not lead to any confusion as the expressions given here are only used in this Appendix. As explained in Sec.~\ref{secquartic} we can ignore the first four terms of $S_1$ in \eqref{4.7} that involve derivatives of $\bar{\Phi}$ as they cannot contribute to the pole part of $\Gamma_4$. (Of course they will make a contribution to the full finite part of the expression for $\Gamma_4$.) We have defined
\begin{subequations}\label{C8}
\begin{align}
\zeta_{1i}(x)&= \zeta_1\,\bar{\Phi}_i(x),\label{C8a}\\
\bar{\zeta}_{1i}(x)&=\zeta_1\,\bar{\Phi}^\dagger_i(x),\label{C8b}\\
\zeta_{2Ai}(x)&= \zeta_2\,\lbrack T_A\bar{\Phi}(x)\rbrack_i,\label{C8c}\\
\Bar{\zeta}_{2Ai}(x)&= \zeta_2\,\lbrack \bar{\Phi}^\dagger(x) T_A\rbrack_i.\label{C8d}
\end{align}
\end{subequations}

If we examine $\langle S_2^2\rangle$ and use \eqref{C1} it is clear that $\langle S_{2\Phi}S_{2G}\rangle=0$ as no connected diagrams can result. For $\langle S_{2\Phi}^2\rangle$ we need equal numbers of $\Phi_i$ and $\Phi^\dagger_j$ terms. It is easily seen that
\begin{equation}
\langle S_{2\Phi}^2\rangle=-\int dv_xdv_{x'}\Big\lbrack 4\,\alpha_{ij}(x)\,\beta_{ij}(x')+\gamma_{ij}(x)\,\gamma_{ji}(x')\Big\rbrack\Delta^2(x,x').\label{C9}
\end{equation}
For $\langle S_{2G}^2\rangle$ we find
\begin{align}
\langle S_{2G}^2\rangle&=-\int dv_xdv_{x'}\Big\lbrack 2\,\epsilon(x)\epsilon(x')+4\,e_s^2\,g^2\,\tau_A(x)\tau_A(x') + 2\,g^4\,\rho_{AB}(x)\rho_{AB}(x')\Big\rbrack\nonumber\\
&\qquad\qquad\qquad\times\, \widetilde{G}^{\mu\nu}(x,x')\widetilde{G}_{\mu\nu}(x,x').\label{C10}
\end{align}
If we now use \eqref{B6} and \eqref{B7} we obtain 
\begin{equation}
{\textrm{PP}}\left\lbrace\langle S_{2}^2\rangle\right\rbrace=\frac{i}{8\pi^2\epsilon}\int dv_x\Big\lbrack 4\,\alpha_{ij}\,\beta_{ij}+\gamma_{ij}\,\gamma_{ji}+6\,\epsilon^2+12\,e_s^2\,g^2\,\tau^2_A + 6\,g^4\,\rho^2_{AB}\Big\rbrack.\label{C11}
\end{equation}

For $\langle S_{1}^2S_2\rangle$ we expand using \eqref{C1} and \eqref{C6}. This gives rise to twenty separate terms some of which vanish because of the pairing relations on the quantum fields given in Sec.~\ref{secBose}. We leave out all of these intermediate details for brevity. The net result is that
\begin{align}
{\textrm{PP}}\left\lbrace\langle S_{1}^2S_2\rangle\right\rbrace&=\frac{i}{8\pi^2\epsilon}\int dv_x\Big\lbrack 2\,(\zeta_{1i}\zeta_{1j}+ \zeta_{2Ai}\zeta_{2Aj})\alpha_{ij} + 2\,(\bar{\zeta}_{1i}\bar{\zeta}_{1j} + \bar{\zeta}_{2Ai}\bar{\zeta}_{2Aj} )\beta_{ij}\nonumber\\
&\quad-2\,({\zeta}_{1i}\bar{\zeta}_{1j}+{\zeta}_{2Ai}\bar{\zeta}_{2Aj})\gamma_{ji} -4\,\epsilon\,\zeta_{1i}\bar{\zeta}_{1i}-4\,e_s\,g(\zeta_{1i}\bar{\zeta}_{2Ai}+ \bar{\zeta}_{1i}{\zeta}_{2Ai})\tau_A\nonumber\\
&\quad -4\,g^2\,\zeta_{2Ai}\bar{\zeta}_{2Bi}\rho_{AB}\Big\rbrack.\label{C12}
\end{align}

Finally, for $\langle S_{1}^4\rangle$ it is straightforward to show from \eqref{C6} and \eqref{C7} that
\begin{align}
{\textrm{PP}}\left\lbrace\langle S_{1}^4\rangle\right\rbrace&=-\,\frac{3\,i}{4\pi^2\epsilon}\int dv_x\Big\lbrack 4\,(\zeta_{1i}\bar{\zeta}_{1i})^2+ (\zeta_{2Ai}\bar{\zeta}_{2Bi}+\bar{\zeta}_{2Ai}\zeta_{2Bi})^2\nonumber\\
&\qquad + 2\,({\zeta}_{1i}\bar{\zeta}_{2Ai} + \bar{\zeta}_{1i}{\zeta}_{2Ai} )^2\Big\rbrack.\label{C13}
\end{align}

It now remains to eliminate the auxiliary expressions that we introduced in \eqref{C4}, \eqref{C8}, \eqref{3.8a}, \eqref{3.8}, and \eqref{4.9}. Using \eqref{4.9} it can be shown that \eqref{C11} becomes
\begin{equation}
{\textrm{PP}}\left\lbrace\langle S_{2}^2\rangle\right\rbrace = \frac{i}{8\pi^2\epsilon}\int dv_x \Big( A_1\,|\bar{\Phi}|^4+B_1\,\tau_A^2+D_1\,\rho_{AB}^2\Big),\label{C14}
\end{equation}
where we have defined
\begin{subequations}\label{C15}
\begin{align}
A_1&=6\,e_s^4+2\,\zeta_1^2(\zeta_1-2\,e_s)^2-\frac{2}{3}\,\lambda\,\zeta_1(\zeta_1-2\,e_s)\nonumber\\
&\quad - \frac{2}{3}\,\lambda\,C_2\,\zeta_2(\zeta_2-2\,g)+\frac{1}{9}(d_R+4)\,\lambda^2,\label{C15a}\\
B_1&=12\,e_s^2\,g^2+4\,\zeta_1\,\zeta_2\,(\zeta_1-2\,e_s)(\zeta_2-2\,g),\label{C15b}\\
D_1&=6\,g^4+2\,\zeta_2^2\,(\zeta_2-2\,g)^2.\label{C15c}
\end{align}
\end{subequations}
It is possible to write \eqref{C14} solely in terms of $|\bar{\Phi}|^4$ by making use of \eqref{A10} and \eqref{A11} but we will defer this until we have the complete expression for ${\textrm{PP}}\lbrace\Gamma_4\rbrace$.

We can write ${\textrm{PP}}\lbrace\langle S_{1}^2S_2\rangle\rbrace$ given in \eqref{C12} in a form analogous to that in \eqref{C14}:
\begin{equation}
{\textrm{PP}}\left\lbrace\langle S_{1}^2S_2\rangle\right\rbrace = \frac{1}{8\pi^2\epsilon}\int dv_x \Big( A_2\,|\bar{\Phi}|^4+B_2\,\tau_A^2+D_2\,\rho_{AB}^2\Big),\label{C16}
\end{equation}
where we have defined
\begin{subequations}\label{C17}
\begin{align}
A_2&=-4\,(\zeta_1-e_s)^2+\frac{2}{3}\,\lambda(\zeta_1^2+C_2\,\zeta_2^2),\label{C17a}\\
B_2&=-8\,\zeta_1\,\zeta_2\,(\zeta_1-e_s)(\zeta_2-g),\label{C17b}\\
D_2&=-4\,\zeta_2^2\,(\zeta_2-g)^2.\label{C17c}
\end{align}
\end{subequations}
(Note that there is great simplification in the Landau-DeWitt gauge \eqref{3.5}.)

Using \eqref{C8} it can be shown that \eqref{C13} becomes
\begin{equation}
{\textrm{PP}}\left\lbrace\langle S_{1}^4\rangle\right\rbrace=-\,\frac{3\,i}{4\pi^2\epsilon}\int dv_x\Big( 4\,\zeta_{1}^4\,|\bar{\Phi}|^4 + 8\,\zeta_1^2\,\zeta_2^2\,\tau_A^2+4\,\zeta_2^4\,\rho_{AB}^2\Big).\label{C18}
\end{equation}

This completes the evaluation of the expressions needed in Sec.~\ref{secquartic}.


\end{document}